\documentclass[a4paper]{PoS}
\usepackage[utf8x]{inputenc}
\usepackage{amsmath,amssymb}
\usepackage{bm}
\usepackage{comment}
\usepackage{graphics,graphicx}
\usepackage[dvipsnames]{xcolor}
\usepackage{slashed}
\usepackage{url}
\usepackage{multicol}

\usepackage{floatrow}
\newfloatcommand{capbtabbox}{table}[][\FBwidth]

\let\OLDthebibliography\thebibliography
\renewcommand\thebibliography[1]{
  \OLDthebibliography{#1}
  \fontsize{7.8125pt}{9pt}
  \setlength{\parskip}{0pt}
  \setlength{\itemsep}{0pt plus 0.5ex}
  
}

\newcommand{\M}{\ensuremath{\mathcal{M}}}
\newcommand{\deltaG}{\ensuremath{\delta\mathcal{G}^V}}
\renewcommand{\O}{\ensuremath{\mathcal{O}}}

\providecommand{\repositoryInformationSetup}{} 
\repositoryInformationSetup

\usepackage{xspace}
\usepackage{bbm}

\newcommand{\Tabref}[1]{Table~\ref{tab:#1}\xspace}

\newcommand{\Figref}[1]{Figure~\ref{fig:#1}\xspace}
\renewcommand{\eqref}[1]{(\ref{eq:#1})\xspace}

\newcommand{\goesto}{\ensuremath{\rightarrow}}

\newcommand{\inverse}{\ensuremath{^{-1}}}                                       

\newcommand{\transpose}{\ensuremath{{}^{\top}}}
\providecommand{\inverse}{\ensuremath{^{-1}}}                                       

\providecommand{\transpose}{\ensuremath{{}^{\top}}}

\let\builtinLaTeX\LaTeX
\def\LaTeX{\builtinLaTeX\xspace}

\title{Progress in Multibaryon Spectroscopy}

\ShortTitle{Multibaryon Spectroscopy}

\author{\speaker{Evan Berkowitz}\thanks{For concision and to foster the notion that this series of symposia forms an ongoing community dialogue I have made an effort to cite, where possible, work from these and prior proceedings in favor of work published elsewhere.  Slides are available at \mbox{\protect\url{https://indico.fnal.gov/event/15949/session/2/contribution/314}}.
This is document RBRC-1299.
}\\
        Institut f\"{u}r Kernphysik and Institute for Advanced Simulation, Forschungszentrum J\"{u}lich\\
        E-mail: \email{e.berkowitz@fz-juelich.de}}

\author{David Brantley, Kenneth McElvain, and Andr\'{e} Walker-Loud \\ Nuclear Science Division, Lawrence Berkeley National Laboratory \\ \email{dbrantley@lbl.gov}, \email{KSMcElvain@lbl.gov}, \email{AWalker-Loud@lbl.gov}}
\author{Chia Cheng Chang \\ Interdisciplinary Theoretical and Mathematical Sciences Program (iTHEMS) RIKEN \\ \email{ChiaChang@lbl.gov}}
\author{M. A. Clark \\ NVIDIA Corporation \\ \email{mclark@nvidia.com}}
\author{Thorsten Kurth \\ National Energy Research Scientific Computing Center, Lawrence Berkeley National Laboratory \\ \email{tkurth@lbl.gov}}
\author{B\'{a}lint Jo\'{o} \\ Jefferson Lab \\ \email{bjoo@jlab.org}}
\author{Henry Monge-Camacho and Amy Nicholson \\ Department of Physics and Astronomy, University of North Carolina \\ \email{hjmonge@email.unc.edu}, \email{annichol@email.unc.edu}}
\author{Enrico Rinaldi \\ RIKEN BNL Research Center \\ \email{erinaldi@bnl.gov}}
\author{Pavlos Vranas \\ Nuclear and Chemical Sciences Division, Lawrence Livermore National Laboratory \\ \email{vranas2@llnl.gov}}

\abstract{Anchoring the nuclear interaction in QCD is a long-outstanding problem in nuclear physics. While the lattice community has made enormous progress in mesonic physics and single nucleon physics, continuum-limit physical-point multi-nucleon physics has remained out of reach. I will review CalLat's strategy for multi-nucleon spectroscopy and our latest results.}

\FullConference{The 36th Annual International Symposium on Lattice Field Theory - LATTICE2018\\
        22-28 July, 2018\\
        Michigan State University, East Lansing, Michigan, USA.}

\begin{document}
\maketitle

\section{Introduction}

Nuclear physics remains quantitatively disconnected from the input parameters of the Standard Model.
A long-outstanding dream is to determine the nuclear interaction directly from the fundamental strong interaction of quarks and gluons\cite{Savage:2005ma,Aoki:2007zf,Beane:2008ia,Detmold:2009zz,Hatsuda:2014kma,Orginos:2011zz,Doi:2012ab,Walker-Loud:2014iea,Yamazaki:2015nka,Savage:2016egr,Davoudi:2017ddj}.
The difficulty, of course, is that QCD, the theory of those particles' interactions is analytically intractable at the low energies of nuclear physics and nonperturbative methods are required.
The only known method without uncontrollable uncertainties is lattice QCD.

While the experimental coverage of two-nucleon scattering processes is extensive and nuclei with more than two baryons and nuclear reactions are well-measured, a quantitative understanding of nuclear physics directly from QCD remains desirable.
Three obvious reasons are, first, to extend the reach of the explanatory power of the Standard Model; second, to enable precision tests of the Standard Model; third, to understand matter in extreme conditions.

Without an understanding of what the Standard Model predicts, it is impossible to quantify any potentially-observed new physics in low-energy nuclear experiments.
For example, a complete understanding of any experimental signal of neutrinoless double beta decay requires quantifying how the nucleon axial coupling is modified in a nuclear environment and quantifying how short-distance beyond-the-Standard-Model operators infect the nuclear interaction.

The lattice has long held out the promise of precision non-perturbative QCD calculations.
In recent years we have seen the development of continuum-limit, physical-point calculations.
In particular, the spectrum of single hadrons is well-determined and dynamical electromagnetism accounted for.

Multi-particle phenomena are much less constrained.
Those cases with the greatest precision are entirely mesonic.
At first glance, the mesonic case is simpler because the number of required Wick contractions is smaller.
On the other hand, many channels have quark-line disconnected diagrams that require stochastic methods.
More seriously, the baryon signal-to-noise problem currently encumbers attempts to reproduce the precision of mesonic scattering, and only worsens with baryon number.

Nevertheless, a variety of calculations studying the few-baryon sector exist\cite{Beane:2006mx,Ishii:2006ec,Beane:2010hg,Inoue:2010hs,Inoue:2010es,Nemura:2008sp,Yamazaki:2010ya,Beane:2011iw,Doi:2011gq,Inoue:2011ai,Yamazaki:2011nd,Aoki:2012tk,Beane:2012ey,Beane:2012vq,Yamazaki:2012hi,Yamazaki:2012fn,Beane:2013br,Murano:2013xxa,Orginos:2015aya,Yamazaki:2015asa,Yamazaki:2015nka,Berkowitz:2015eaa,Wagman:2017tmp}.
In the remainder of these proceedings, I will review the methods and calculations of the CalLat collaboration.
\section{Applying the L\"{u}scher Finite-Volume Formalism}

The L\"{u}scher finite-volume formalism is the method by which finite-volume energy levels can be transformed into information about infinite-volume scattering\cite{luscher:1986n1,luscher:1986n2,Wiese:1989,Luscher:1991n1,Luscher:1991n2,Rummukainen:1995vs,Feng:2004ua,He:2005ey,Kim:2005gf,Lage:2009zv,Bernard:2010fp,Fu:2011xz,Bernard:2012bi,Briceno:2012yi,Gockeler:2012yj,Guo:2012hv,Briceno:2014oea,Lee:2017igf,Briceno:2018bnl,Davoudi:2017ddj,Brett:2017yhm,Bulava:2017stw}.
It is quite general and is applied with enormous success in the meson sector.
To extract infinite-volume scattering data one must solve the determinant equation
\begin{equation}\label{eq:luescher}
    \det\left[\left(\M\right)^{-1} + \deltaG\right] = 0,
\end{equation}
the quantization condition that the interacting states exactly fit into the box, under the assumption that the box is substantially larger than the range of the interaction.
The matrix $\deltaG$ contains information about the finite-volume spectrum and the finite volume itself, while $\M$ encodes the infinite-volume scattering amplitudes of interest at the respective energies.
By changing the size of the box or boosting the center of mass we can access different kinematic points and build a map of the scattering data as a function of momentum.
The formal development of the L\"{u}scher method is quite advanced, with an understanding of how to handle inelasticities and coupled channels\cite{Lage:2009zv,Wilson:2016rid}, twisted boundary conditions\cite{Briceno:2013hya,Korber:2015rce}, and other complications and progress towards an understanding of the three-body sector advancing rapidly\cite{Hansen:2013dla,Hansen:2014lya,Hansen:2015azg,Briceno:2016ffu,sharpe:2018
}.

The nuclear force is not a central interaction, and orbital and spin angular momenta $L$ and $S$ are not conserved separately.
Instead, only their total $J$ is conserved.
However, finite-volume energy eigenfunctions must satisfy the quantization condition, which arises from the boundary conditions of the finite volume.
The states therefore have the symmetry of the finite volume, rather than the SO(3) symmetry expected from the conservation of angular momentum in an infinite volume.
In a cubic box, the states therefore fall into irreducible representations of the octahedral group $O_H$.
Once the center of mass is boosted, relativistic length contraction further breaks this symmetry and the irreps split into their descendants in the remaining symmetry group.
The reduced symmetry implies that finite-volume eigenstates will be superpositions of infinite-volume channels that would not otherwise mix.
We say that the infinite-volume states with good angular momentum quantum numbers are \emph{subduced} onto states labeled by $\mu$ in irreps labeled by $\Lambda$.

The required energy levels may be extracted by any reliable spectroscopic method.
In lattice QCD we typically extract the spectrum by fitting Euclidean-time correlation functions.
Correlation functions,
\begin{equation}
    C^{ij}(t) = \left\langle \Omega \middle |
    \O^{i\ [J'\ell'S']}_{\Lambda'\mu',I'm'_I}(t)
    \O^{j\ [J\ell S]\dagger}_{\Lambda\mu Im_I}(0)
    \middle| \Omega \right\rangle
\end{equation}
where $i$ and $j$ run over some set of interpolating operators with the quantum numbers of interest and the primed indices label quantities at the sink and unprimed indices quantitites at the source, admit a spectral decomposition
\begin{equation}
    C^{ij}(t) = \sum_n z_n^i z_n^{j\dagger} e^{-E_n t}
\end{equation}
where the sum is over eigenstates with the quantum numbers of interest, $E_n$ the energy of the $n^\text{th}$ state in the spectrum, $t$ the Euclidean time, and $z_n$ are overlap factors that express how well the interpolating operator gives the $n^\text{th}$ state.

In the long-time limit the correlator is dominated by the lowest-lying state, and one can extract the energy by calculating the effective mass,
\begin{align}
    m(t) &= - \partial_t \ln C(t)
    &
    E_0 &= \lim_{t\goesto\infty} m(t)
\end{align}
where the derivative is typically computed as a finite difference between timeslices on the lattice.
Other more sophisticated spectroscopic techniques are available.

Alternatives to the L\"{u}scher formalism include the the potential method and the unitary isobar formalism.
An enormous amount of progress both formal and numerical has been achieved, and new results and updates appear elsewhere in these proceedings\cite{mai:2018,ma:2018,yamazaki:2018,namekawa:2018,iritani:2018,doi:2018}.
Another approach is to put an EFT of interest into a finite volume and tune its low-energy constants to match the spectrum directly, rather than converting the energy levels to infinite-volume observables\cite{guo:2018}.
The approach taken in this work is the application of L\"{u}scher's formalism.

In baryonic channels the signal-to-noise problem prevents a clean examination of the late-time limit.
While new methods, such as phase reweighting\cite{Wagman:2017jva}, have appeared, their application so far has remained limited.
One hope is to find interpolating operators which have very little excited state contamination, so that they plateau early before the noise overwhelms the signal.
In the next two sections we discuss the construction of interpolating operators and what may be done to extend plateaus earlier in Euclidean time.

\section{Single-Nucleon Improvements}

Rather than fight the noise, one may try to remove excited state contamination to produce an early, reliable plateau, by taking linear combinations of interpolating operators.
One worrisome source of excited state contamination comes from the inelastic relaxation of the nucleon to its ground state.
By improving the individual nucleons and using those nucleons in a multi-nucleon calculation, we can get a better handle on the elastic processes of interest.

A large basis of operators for nucleons has long been known\cite{Basak:2005ir} which each give different overlap with the ground and tower of excited states.
Since we are only interested in the scattering of nucleons and not any inelastic excited states, we can take the simpler approach of using a point-like operator and a similar, smeared operator.
By changing the smearing we can get an operator with markedly different overlap factors from the point operator.
By finding a good linear combination of the point and smeared operator we improve the ground state, throwing much of the tower of excited states into the garbage pail of the orthogonal linear combination.

In a truly variational method, the same improvement would be done at the source and the sink in a positive-definite way.
However, the high cost of multinucleon contractions has led us to use fixed source interpolators and only improve the sink.
To get a procedural improvement of the single nucleon, rather than a so-called `$\chi$-by-eye' improvement, we apply the Matrix Prony method\cite{Beane:2009kya,Orginos:2014pma}, which we summarize here.
We hope to extract the ground state of some vector of correlators $y(t)$ which should all have the same ground state.
These $y$ might be generated, for example, by using a fixed smeared source and both point and smeared sinks, as we do here.
We assume there exists a transfer operator $\hat{T}(\tau)$ such that
\begin{equation}
    y(t+\tau) = \hat{T}(\tau)y(t).
\end{equation}
By multiplying both sides on the right by $y\transpose$ and assuming $\hat{T} = M\inverse V$ we can solve via
\begin{align}
    M &= \left[ \sum_{t=t_i}^{t_f}y(t+\tau)y\transpose(t)\right]\inverse
    &
    V &= \left[ \sum_{t=t_i}^{t_f}y(t)y\transpose(t)\right]\inverse
\end{align}
under the assumption that only as many excited states as entries in $y$ contribute from times $t_i$ to $t_f$.
Knowing $M$ and $V$ then allow us to construct $\hat{T}$.
Diagonalizing $\hat{T}$ gives orthogonal linear combinations of $y$ with an isolated exponential decay.
We observe that the lower-energy state is typically reliable while the orthogonal higher-energy state is typically much more noisy and tends to capture violations of the few-excited-states assumption.

Taking the linear combination that corresponds to the lowest energy state allows us to construct what we call `calm baryons', as they are less excited\cite{Berkowitz:2017smo}.
With a reliable linear combination the calm baryon correlator plateaus substantially earlier (see, for example, Fig.~1 of Ref.~\cite{Berkowitz:2017smo}).
These linear combinations can be naturally incorporated into baryon block methods\cite{Doi:2012xd,Detmold:2012eu,Orginos:2012js,Vachaspati:2014bda}.
For a single nucleon this doesn't yield any substantive computational advantage, but in the multinucleon case it reduces substantially the number of contractions needed to eliminate these single-nucleon inelastic excited states.
More importantly, incorporating these calm baryons into multinucleon calculations dramatically reduces the excited state contamination, as can be see in Fig.~3 of Ref.~\cite{Berkowitz:2017smo}.
\section{Spatially-Displaced Two-Nucleon Operators}

The analysis of correlation functions $C^{ij}(t)$ can pose a nontrivial difficulty.

Because they are substantially cheaper in terms of contractions, mesonic calculations have had enormous success applying a variational method, where good combinations of interpolating operators yield very flat effective masses $m(t)$.
To be positive-definite, these variational methods require the same set of operators at both source and sink.

The high cost of constructing a complete cross-correlator from a large set of baryonic interpolators has held back calculations with baryons, which have only recently begun applying any truly positive-definite variational method\cite{Francis:2018qch}.

The variational method relies on interpolating operators projected to definite center-of-mass and definite irrep.
In other words, the noninteracting states needed are superpositions of states with baryons of definite momenta.
With point-to-all propagators we retain the ability to construct baryons with definite momenta at the sink, but the source is more problematic.
While it is easy to project quark sources to a definite momentum, without a combinatorial explosion it's not simple to project a baryon source to a definite momentum.
Without an all-to-all method we're stuck considering correlators where the source consists of baryons with definite spatial location and the sink consists of baryons with definite momenta and thus are not positive-definite.
This indefinite correlation can cause problems of false plateaus\cite{Aoki:2017byw}.

Schematically, such a correlator looks like
\begin{equation}
    \left \langle \Omega \middle |
    \O(\vec{P}, \vec{p};\ t) \O^{\dagger}(\Delta \vec{x};\ 0)
    \middle| \Omega \right \rangle
\end{equation}
where $\vec{P}$ is the total center of mass momentum, $\vec{p}$ the relative momentum, $t$ the Euclidean time, and $\Delta \vec{x}$ a triplet of integers giving the relative spatial displacement of two baryons.
The operators $\O$ themselves are built from two single-baryon operators.

By combining different relative momenta $\vec{p}$ we can project the sink to a noninteracting eigenstate.
After selecting an irrep $\Lambda$ one then can select an $n^2$, where $\vec{n}$ is a triplet of integers that corresponds to the momentum $\vec{p}$.
Increasing $n^2$ increases the energy of the target noninteracting state.
A comprehensive set of states of good cubic symmetry was tabulated by Luu and Savage\cite{Luu:2011ep}\footnote{The \href{http://www.arxiv.org/abs/1101.3347}{arXiv:1101.3347 version} contains substantially more tables and useful information, which the referee requested removed in the published version, according to private correspondence.}.

\begin{figure}[htbp]
    \centering
        \includegraphics[width=0.45\textwidth]{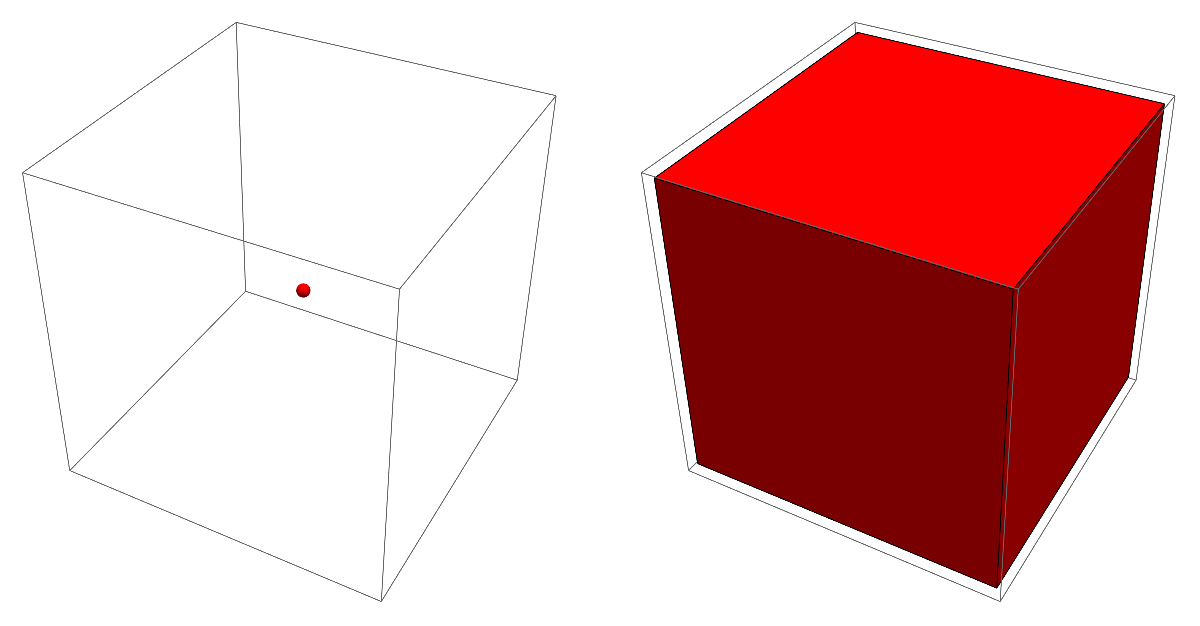}
        \includegraphics[width=0.45\textwidth]{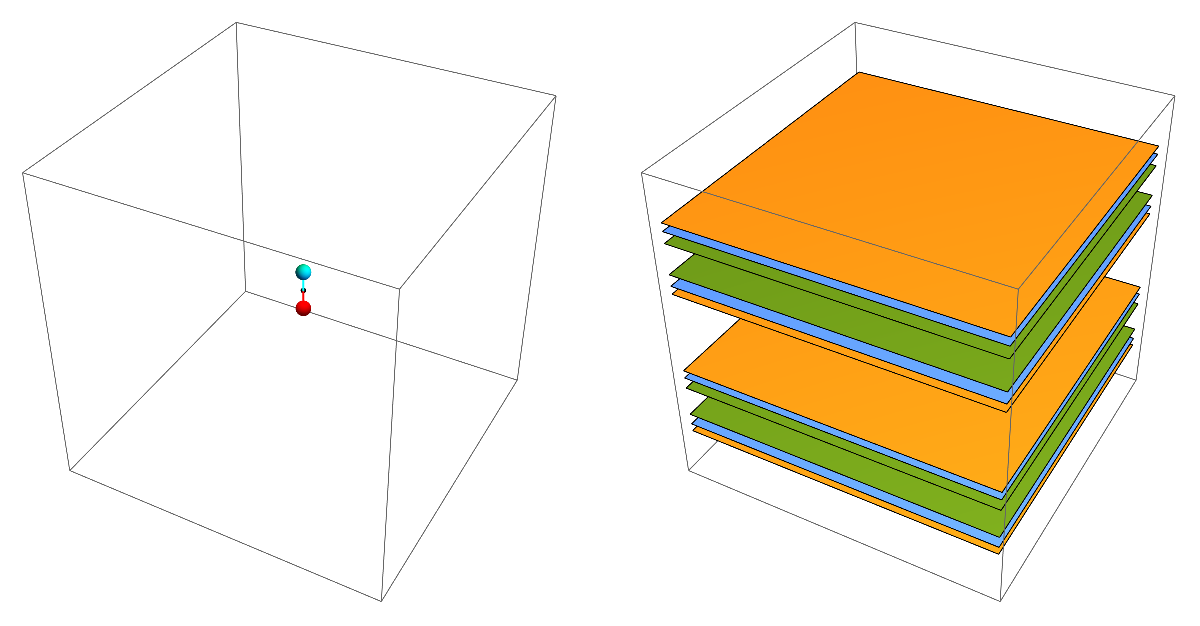}
        \\
        \includegraphics[width=0.45\textwidth]{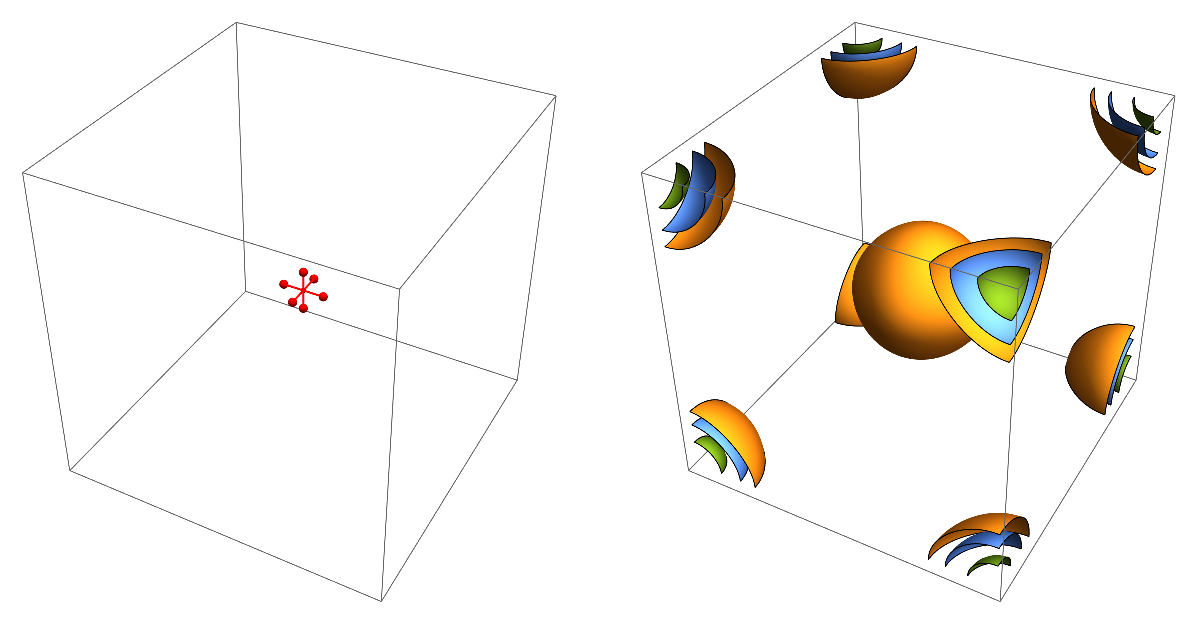}
        \includegraphics[width=0.45\textwidth]{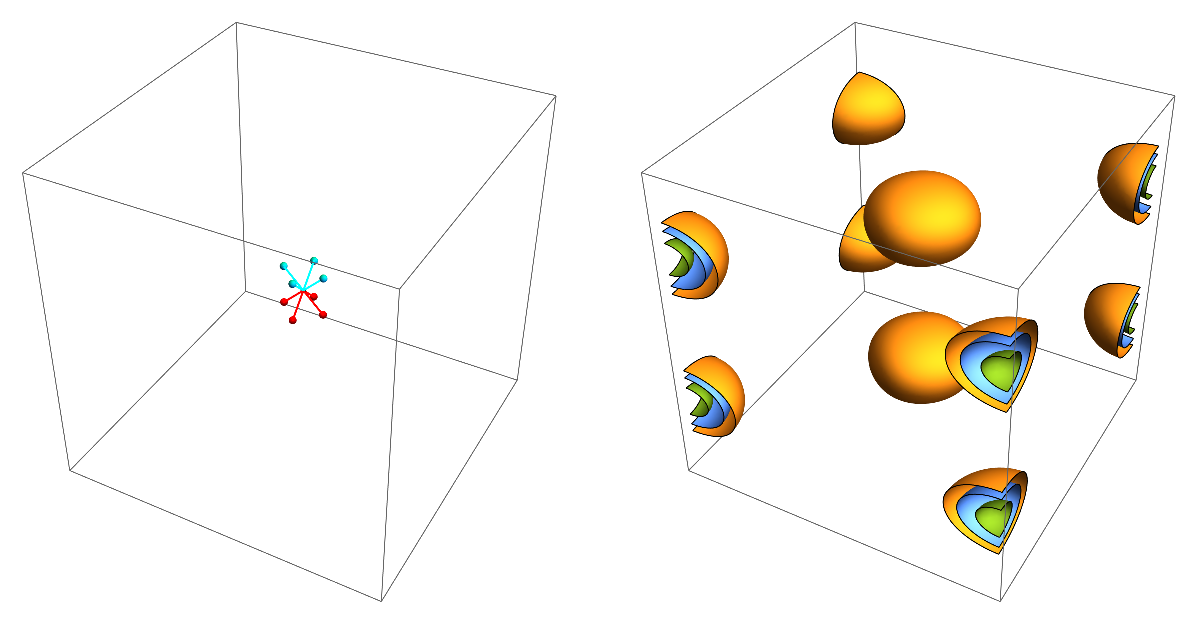}
    \caption{The first and third columns show two of the lowest $n^2$ states in the $A_1^+$ and $T_1^-$ irreps, respectively, in momentum space---red (blue) indicates relative momenta that should be added (subtracted).  The second and fourth columns show some contours of constant square magnitude of their fourier transforms into position space---contours of low square magnitude are omitted for visual clarity.}
    \label{fig:irreps}
\end{figure}

\begin{figure}
\begin{floatrow}
    \capbtabbox{%
        \begin{tabular}{cccl}
        $\Delta\vec{x} \sim $   &        name    &      \# 
                                                                    &   geometric shape                 \\
        \hline
        (0,0,0)                 &        local   &      1           &   point                           \\
        (0,0,1)                 &        {\color[rgb]{0,0.666666,0}face}    &      6           &   octahedron                      \\
        (0,1,1)                 &        {\color[rgb]{0,0,1}edge}    &      12          &   cuboctahedron                   \\
        (1,1,1)                 &        {\color[rgb]{1,0,0}corner}  &      8           &   cube                            \\
        (0,1,2)                 &                &      24          &   truncated octahedron            \\
        (1,1,2)                 &                &      24          &   (small) rhombicuboctahedron     \\
        (1,2,3)                 &                &      48          &   great rhombicuboctahedron       \\
        \end{tabular}
    }{%
        \caption{Displacements $\Delta \vec{x}$ corresponding to different spatial geometries, their corresponding names, costs, and solids.
            The displacements are meant to indicate values which are 0 or otherwise equal.
            For example (3,3,4) has the same properties as (1,1,2) while (4,4,4) has the same properties as (1,1,1).
            The great rhombicuboctahedron is also called the truncated cuboctahedron.
                    \label{tab:sources}}%
    }
    \ffigbox[0.3\textwidth]{
        \includegraphics[width=0.3\textwidth]{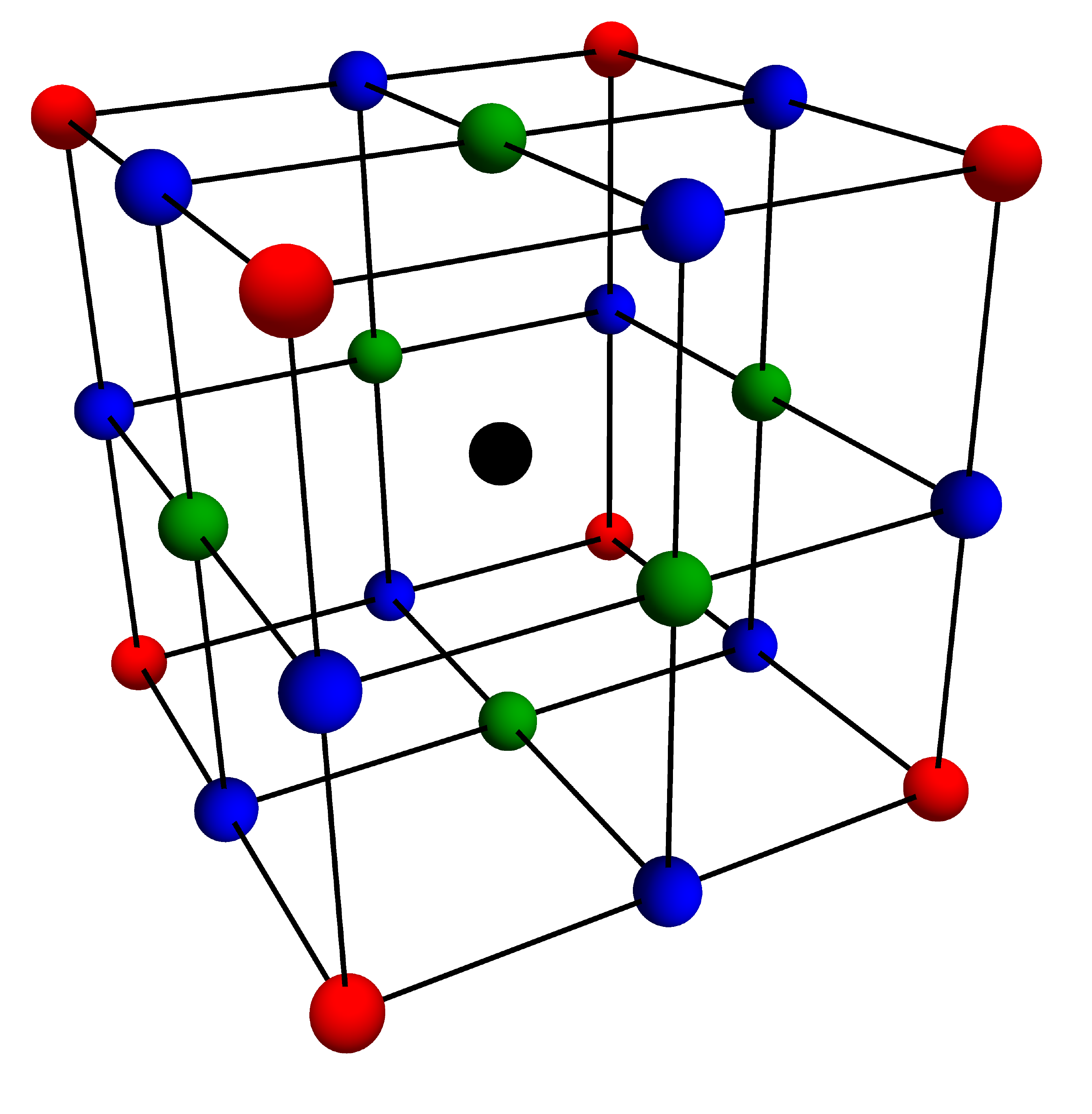}
    }{
        \caption{The cubic rotations of (0,0,1), (0,1,1), and (1,1,1) displacements land on the {\color[rgb]{0,0.666666,0}faces}, {\color[rgb]{0,0,1}edges}, and {\color[rgb]{1,0,0}corners} of a cube around the origin, respectively.\label{fig:cube}}
    }
\end{floatrow}
\end{figure}

On the source side we can construct operators with good overlap with particular irreps.
In the simplest case $\Delta \vec{x} = 0$ we get overlap with $A_1^+$ of all $n^2$ and no other irrep.
As $(\Delta x)^2$ increases one can construct more and more complicated operators and, therefore, irreps.

Spatial displacements exhibit different types of symmetry under the octahedral group.
The zero displacement, of course, is completely symmetric; to access parity-odd partial waves one must avoid the accidental parity symmetry that occurs when two baryons are created on the same lattice site.
Under the action of the octahedral group a displacement along a lattice direction $\Delta \vec{x} \sim (0,0,1)$ generates a total of six displacements, $(0,1,0)$, $(1,0,0)$, and all their additive inverses.
Displacements proportional to $(0,1,1)$ yield a family of twelve displacements (three rearrangements, each with four possible sign assignments for the nonzero entries).
The different shapes and their respective number of vertices, which corresponds to the cost of constructing a source of that shape are detailed in \Tabref{sources}.
The names we give to different displacements correspond to locations on a cube surrounding the origin, as shown in \Figref{cube}.

By scaling up that cube we spread the nucleons apart.
By using different sources we achieve different overlaps with different cubic irreps.
We have previously studied the agreement between different displacements and face, edge, and corner sources\cite{Nicholson:2015pys,Berkowitz:2015eaa}.
By studying the overlap of faces, edges, and corners onto noninteracting states as a function of cube size we can understand how to build sources that have good overlap with many different channels of interest.

For example, restricting a calculation to a local source gives very good overlap with $A_1^+$ but no overlap with other irreps (it is particularly clear that access to the parity-odd irreps is exactly zero by symmetry).
If we picked the maximal possible displacement, $\Delta \vec{x} = (1,1,1)L/2$, where $L$ is the linear size of the cubic volume, we need only perform one additional solve, as that \emph{antipodal} displacement goes to itself under the action of the octahedral group, once accounting for periodic boundary conditions.
This source too gives overlap with $A_1^+$ but no overlap with parity-odd or other parity-even irreps.

Note that a cube around the origin is also a cube of a different size around the the antipode.
This suggests a cost-saving measure: if we do ten solves---at the origin, the antipode, and on a set of corners---we can build a large variety of sources.
Obviously, we can build 10 point sources, an antipodal source, a corner source around the origin on a cube of size $\Delta x$, a corner source around the antipode on a cube of size $L/2 - \Delta x$.
By combining two solves from a corner source we can construct the displacements for 2 faces, 1 edge, and $\frac{1}{2}$ a corner source on a cube of size $2\Delta x$, albeit not around a common point.
However, as the contraction costs are not trivial, we have not yet taken advantage of these other sources.

By trying to maximize the overlap between our position-space displaced sources and the lowest noninteracting states in a variety of cubic irreps for both the cubic source around the origin (OC) and the cubic source around the antipode (AC) one concludes that $\Delta x = L/8$ (or equivalently $3L/8$) is best.
Picking the symmetric $\Delta x = L/4$ often yields a zero overlap with parity-odd cubic states.

\begin{figure}[t!]
    \centering
        \includegraphics[width=\textwidth]{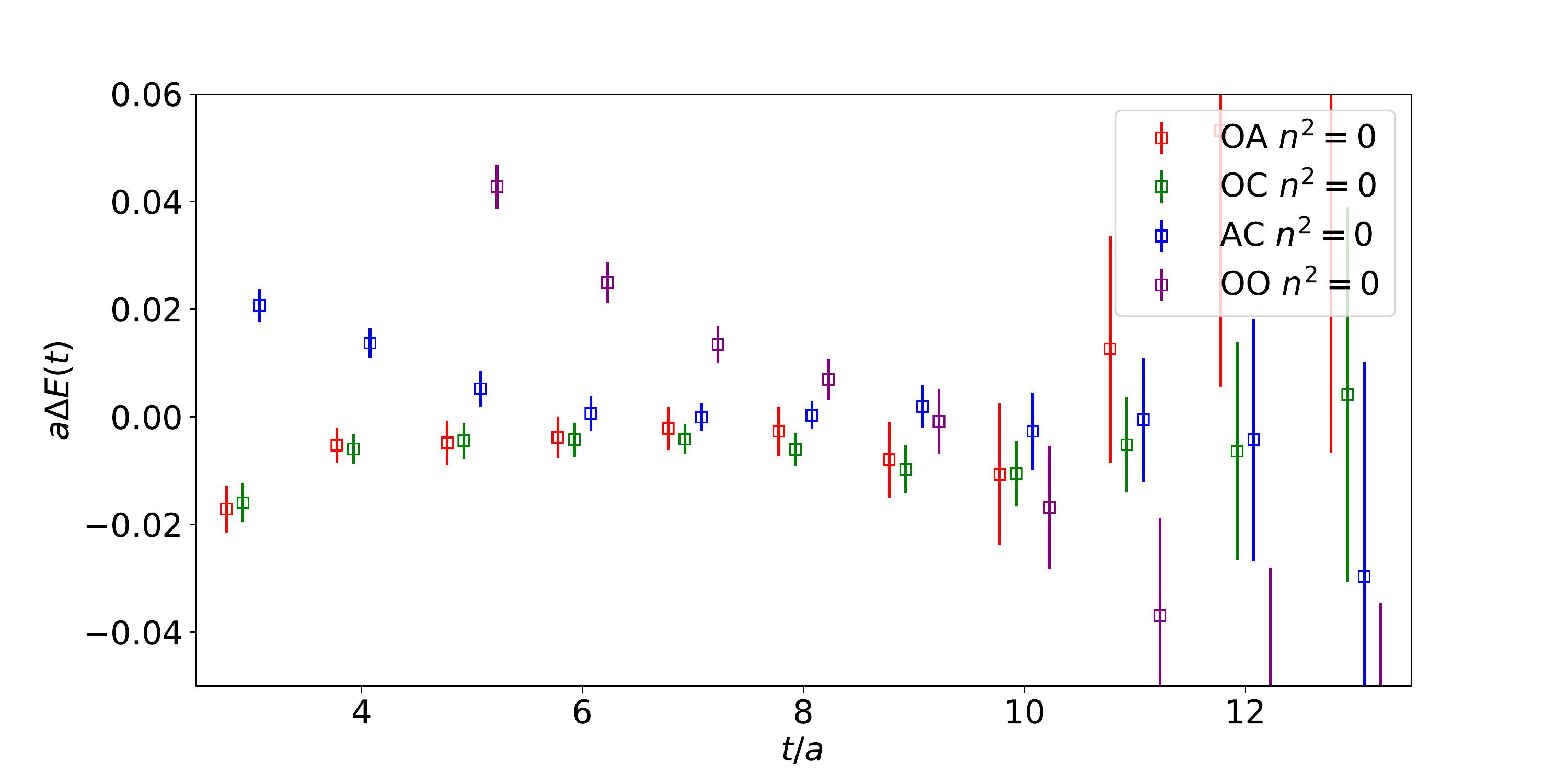} \\
        \includegraphics[width=\textwidth]{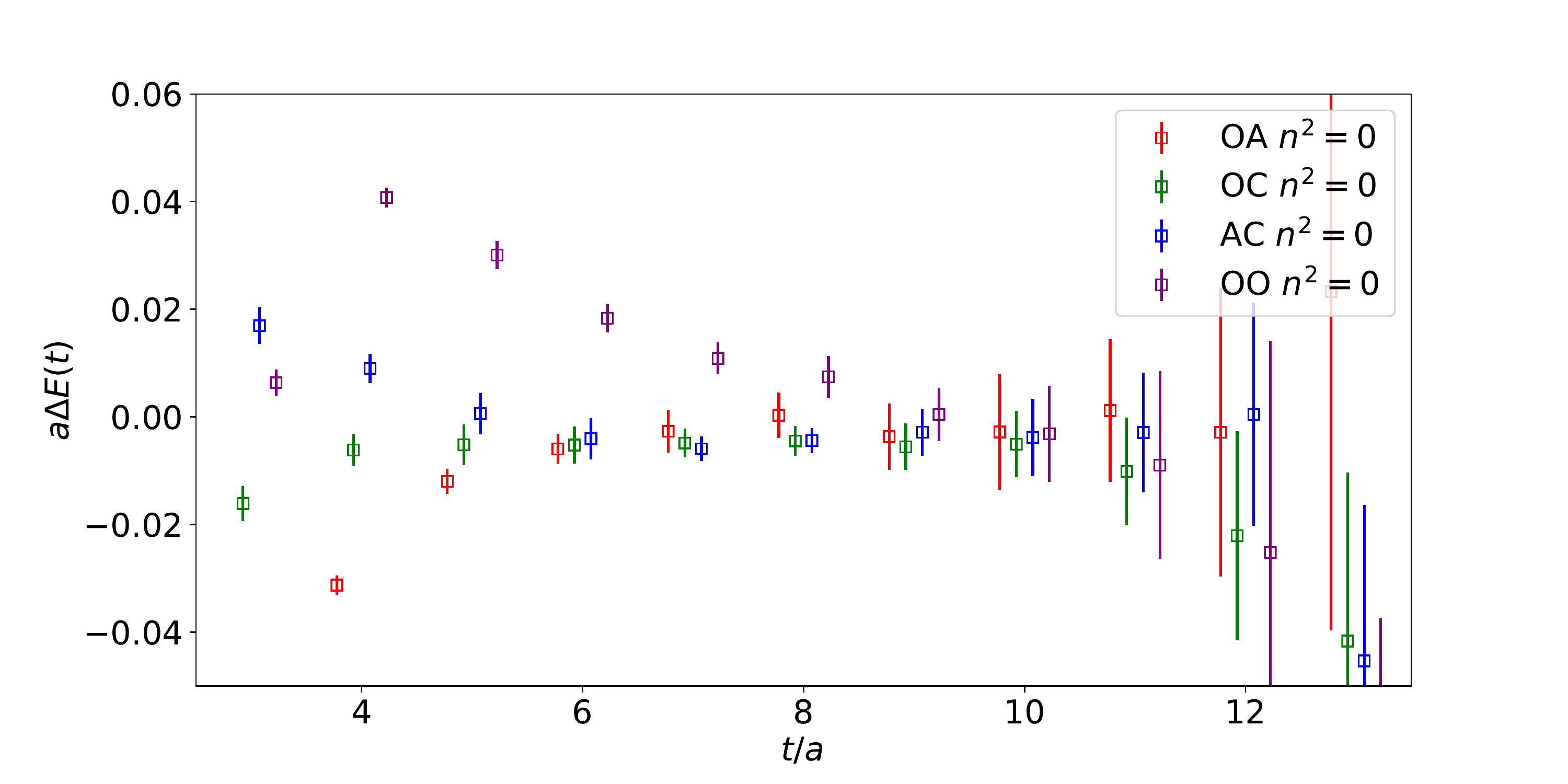}
    \caption{(Top) Four different correlation functions in the isotriplet $A_1^+$ $n^2=0$ channel, which corresponds to the ${}^1S_0$ (dineutron) channel.  (Bottom) Four different correlation functions in the isosinglet $A_1^+$ $n^2=0$ channel, which corresponds to the ${}^3S_1$ (deuteron) channel.  The correlation functions have the same momentum-space sink but different sources---an entirely local source (purple, `OO'), a maximally-displaced source (red, `OA'), a cubic source with $\Delta x = 3L/8$ (green, `OC') and the complementary cubic source with $\Delta x = 1L/8$ (blue, `AC').  The sources are horizontally offset from one another for visual clarity, ordered left-to-right in time in the same order as they are top-to-bottom in the legend.  Calculations were performed on the lattices described in Sec.~\ref{sec:results}.}
    \label{fig:source-study}
\end{figure}

Using those displacements for a $24^3 \times 64$ lattice described in Section~\ref{sec:results} with a calm baryon interpolating operator as previously discussed we can study four different correlation functions in the isotriplet $A_1^+$ $n^2=0$ which corresponds to the ${}^1S_0$ dineutron channel, and the isosinglet $A_1^+$ $n^2=0$ (${}^3S_1$ deuteron) as shown in \Figref{source-study}.
We see that even with calm baryons, the entirely local operator doesn't plateau before hitting the noise.
This may not be surprising, as putting two baryons on top of one another should cause large inelastic distortion, while the tuning of calm baryons kills the excited states of a single nucleon in isolation.

In contrast, the spatially displaced operators have visible, compatible plateaus.
Moreover, the maximally-displaced and OC sources have similar correlation functions while AC has some additional excited state contamination, in comparision.
Again, this is compatible with the idea that calming the baryons is more effective when they are far apart, as AC is on a cube with $\Delta x=L/8\sim0.36$fm while the nucleons in the OC source are more widely separated, $\Delta x = 3L/8\sim1.08$fm.

\section{Preliminary Results at $m_\pi\sim350$MeV}
\label{sec:results}

We generated an ensemble of 10,000 $24^3\times64$ HISQ gauge configurations\cite{Bazavov:2009jc,Bazavov:2009wm,Bazavov:2010pi,Bazavov:2012xda} with a pion mass of about 350 MeV and a lattice spacing of about 0.12 fm ($m_\pi L = 5.1$).
For the valence quark action we use the M\"{o}bius Domain Wall Fermion action described in Ref.~\cite{Berkowitz:2017opd}, with which we previously determined neutrinoless double beta decay matrix elements\cite{Nicholson:2016byl,monge-camacho:2018} and the nucleon axial coupling\cite{Chang:2017oll,Chang:2018uxx} and have been studying the convergence of heavy-baryon $\chi$PT\cite{sallmen:2018}, hadronic CP violation\cite{walker-loud:2018}, methodological improvements\cite{chang:2018,gambhir:2018}, and an independent scale setting\cite{carpenter:2018}.

Mixed action calculations have a long history(see, eg. \cite{Renner:2004ck}); our mixed action has a well-developed EFT (see eg.
 \cite{Bar:2005tu,Tiburzi:2005is,Jiang:2007sn} and references in Ref.~\cite{Berkowitz:2017opd}) and can take advantage of an excellent GPU solver in QUDA\cite{Kim:2014mpa}.
Moreover, the chiral properties of the MDWF valence action helps maintain chiral symmetry in observables.

\begin{figure}[htbp]
    \centering
        \includegraphics[width=\textwidth]{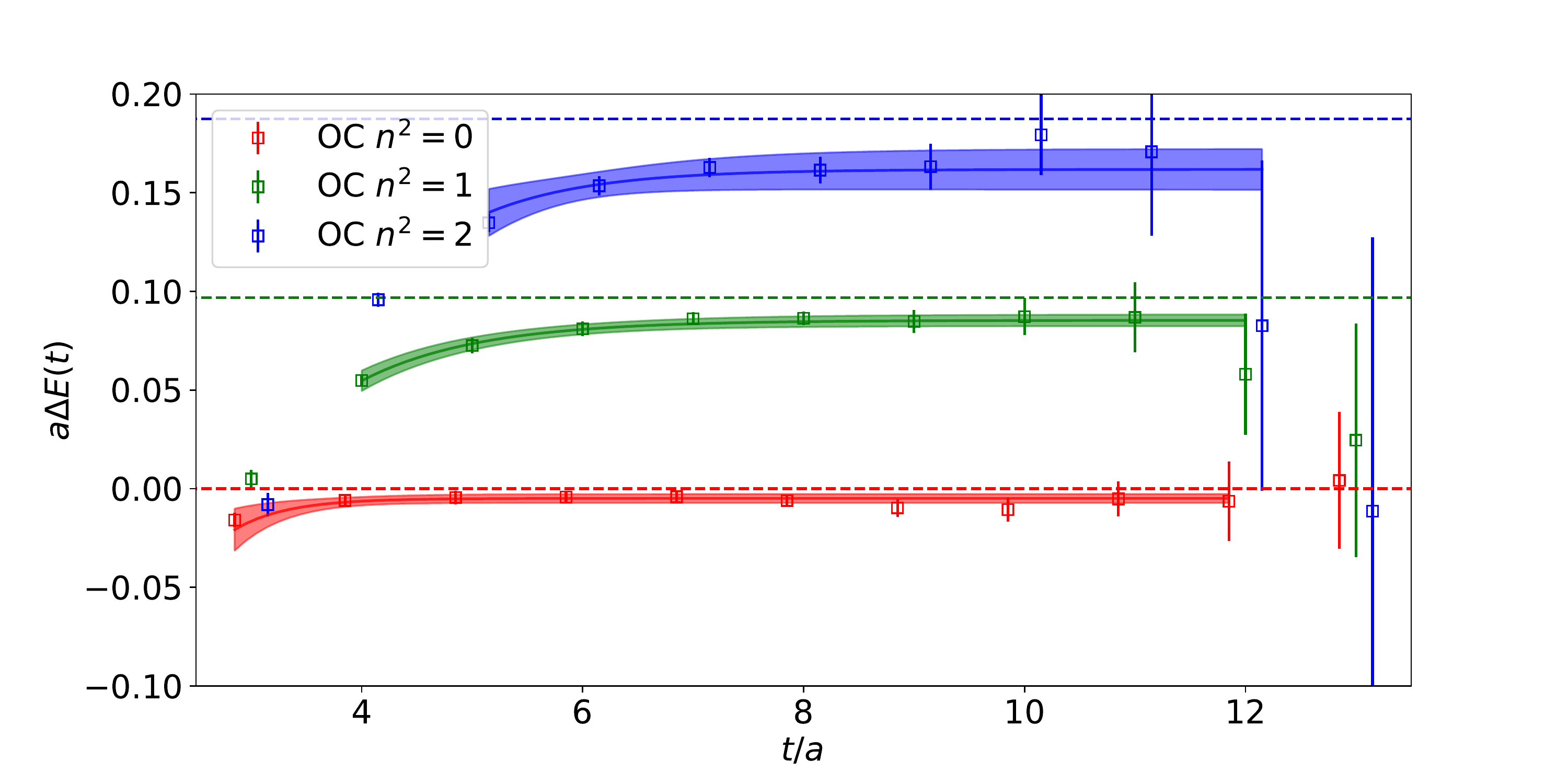} \\
        \includegraphics[width=\textwidth]{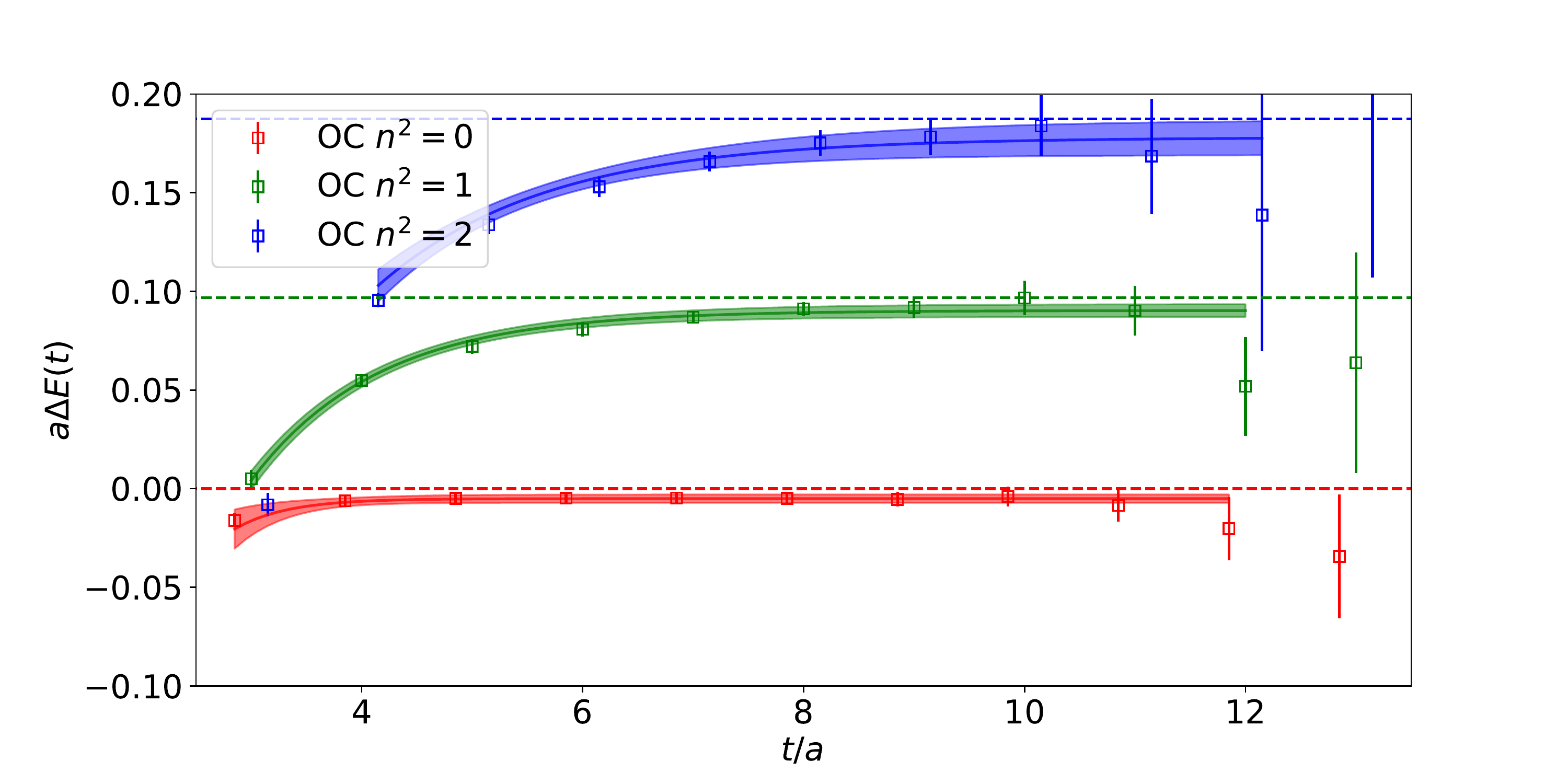} \\
    \caption{The effective masses of $A_1^+$ states for the $I=1$ dineutron (top) and $I=0$ deuteron (bottom) using corner sources and $n^2=0,1,2$ momentum-projected back-to-back calm nucleons as sinks.}
    \label{fig:spectrum}
\end{figure}

\Figref{spectrum} shows effective masses with corner sources and exactly projected to the three lowest-energy $A_1^+$ states for both the isotriplet and isosinglet channels and how they compare to the noninteracting box eigenenergies.
The correlation functions were generated from 1 complete corner source (meaning 9 solves---a center an 8 corners) per configuration and were fit with two exponentials.
The plateau values are reported in \Tabref{plateaus}.

\begin{figure}
\begin{floatrow}
    \capbtabbox{%
        \begin{tabular}{cll}
            $n^2$   &   I=0 [MeV]       &   I=1 [MeV]   \\\hline
            0       &  $-$\phantom{1}$8.3(3.4)$       &  $-$\phantom{1}$8.1(3.7)$    \\
            1       &  $-10.6(5.3)$       &  $-18.8(5.3)$   \\
            2       &  $-16(15)$          &  $-42(17)$
        \end{tabular}
    }{%
        \caption{Energy splittings for $m_\pi = 350$MeV, $a\sim0.12$ fm, $L=2.9$ fm}
        \label{tab:plateaus}
    }%
    \capbtabbox{%
        \begin{tabular}{ccll}
            Ref.                    &  $m_\pi$ [MeV] &     I=0 [MeV]                     &   I=1 [MeV]                   \\\hline
            \cite{Yamazaki:2015asa} &  300           &   $14.5(0.7)(^{+2.4}_{-0.8})$    &  \phantom{1}$8.5(0.7)(^{+1.6}_{0.5})$    \\
            \cite{Orginos:2015aya}  &  450           &   $14.4(^{+3.2}_{-2.6})$         &  $12.5(^{+3.0}_{-5.0})$       
            \\ \phantom{.} 
        \end{tabular}
    }{%
        \caption{Binding energies in two-nucleon LQCD calculations by Yamazaki et al. \cite{Yamazaki:2013rna,Yamazaki:2015vjn,Yamazaki:2015asa} and NPLQCD \cite{Orginos:2015aya} at nearby pion masses.}
        \label{tab:binding-energies}
    }%
\end{floatrow}
\end{figure}

\Tabref{binding-energies} shows binding energies in the isosinglet and isovector channels at two nearby pion masses.
We emphasize that the negative energy splitting we find should not be directly compared to those binding energies---without additional volumes or negative-energy points one cannot determine an infinite-volume binding energy; a crude phase-shift analysis of our results is consistent with no bound state but further study is warranted.
The characteristic lengths $1/\sqrt{\mu B}$ where $\mu$ is the reduced mass and $B$ is the the $n^2=0$ energies are about 2.89 fm, while the box size is about 2.88 fm; such a close match in size may indicate a large finite-volume distortion of the wavefunction.

\section{Outlook}

In addition to the two-nucleon sector, there have been characterizations of the three-baryon and four-baryon sectors.
In the calculations thus far performed the baryons are always created on a single site.
\Figref{source-study} suggests that these calculations may benefit dramatically from a combination of single-nucleon improvement and spatially displaced sources.
As more baryons are added, however, the displaced sources become more expensive.

However, elsewhere in these proceedings the first proof-of-principle calculation of the three-neutron (isospin-$\frac{3}{2}$) system is described\cite{wynen:2018}.
This channel differs from other many-baryon calculations in that the Pauli exclusion principle implies that spatially displaced sources are mandatory, rather than simply beneficial.
To construct spin-$\frac{3}{2}$ interpolators, moreover, each baryon must be on its own site\cite{Luu:2008fg}.
It is natural, then, to try to remove the inelastic single-nucleon excited states in these calculations.

As that calculation matures, the interpretation of the three-body spectrum will present an outstanding theoretical challenge.
The formal development of the three-body formalism is rapidly advancing and connects finite-volume spectra with infinite-volume scattering observables\cite{Hansen:2013dla,Hansen:2014lya,Hansen:2015azg,Briceno:2016ffu,sharpe:2018,mai:2018}.
Rather than extracting the infinite-volume physics in a model-independent way and then fit an EFT of choice to those observables, one may instead put the EFT in a matching finite volume and tune its LECs to match the finite-volume spectrum\cite{guo:2018}---once the LECs are fixed, one may then use the EFT in an infinite volume.
Progress along these lines for analyzing the NPLQCD results at $m_\pi\sim800$MeV with pionless EFT has been reported at the recent Few Body conference\cite{Barnea:2018}, while progress using the Harmonic-Oscillator Basis Effective Theory\cite{McElvain:2016zbm} formulated with periodic boundary conditions was described at the April APS meeting in 2017\cite{McElvain:2017}.

Whatever method used to go from finite to infinite volume, the extraction of a reliable spectrum is paramount.
Ref.\cite{hanlon:2018} details a truly variational calculation in the H-dibaryon channel using distillation at $m_\pi\sim$960, and demonstrates clean, reliable plateaus that begin almost immediately, where a simple point-to-all calculation on the same configurations suffers from bad excited state contamination.
The impressive success of this calculation heralds exciting promise for future progress in the few-baryon sector.
\section*{Acknowledgements}

An award of computer time was provided by the Innovative and Novel Computational Impact on Theory and Experiment (INCITE) program to CalLat (2016) and by the OLCF Director's Discretionary Time (2017).
This research used the NVIDIA GPU-accelerated Titan supercomputer at the Oak Ridge Leadership Computing Facility at the Oak Ridge National Laboratory, which is supported by the Office of Science of the U.S. Department of Energy under Contract No. DE-AC05-00OR22725, and the GPU-enabled Surface and RZHasGPU and BG/Q Vulcan clusters at LLNL.
We thank the LLNL Multiprogrammatic and Institutional Computing program for Grand Challenge allocations on the LLNL supercomputers.

This work was supported by the NVIDIA Corporation (MAC), the DFG and the NSFC Sino-German CRC110 (EB), an LBNL LDRD (AWL), the RIKEN Special Postdoctoral Researcher Program (ER), the U.S. Department of Energy, Office of Science: Office of Nuclear Physics (EB, CCC, TK, HMC, AN, ER, BJ, PV, AWL); Office of Advanced Scientific Computing (EB, BJ, TK, AWL); and the DOE Early Career Award Program (DAB, CCC, HMC, AWL).
This work was performed under the auspices of the U.S. Department of Energy by LLNL under Contract No. DE-AC52-07NA27344 (EB, ER, PV).

Calculations were performed with the \texttt{chroma} software suite and \texttt{quda} solvers built atop the USQCD stack\cite{Edwards:2004sx,Clark:2009wm,Babich:2011np} with the \texttt{HDF5} library\cite{hdf5,Kurth:2015mqa} and managed with \texttt{METAQ} and \texttt{mpi\_jm}\cite{Berkowitz:2017vcp,Berkowitz:2017xna}.
The a12m350 ensemble was generated with the MILC collaboration's  public lattice  gauge  theory code (see \href{http://physics.utah.edu/~detar/milc.html}{http://physics.utah.edu/~detar/milc.html}).

\begin{multicols}{2}
\bibliographystyle{JHEP}
\bibliography{master}

\providecommand{\href}[2]{#2}\begingroup\raggedright\begin{thebibliography}{100}

\bibitem{Savage:2005ma}
M.~J. Savage, \emph{{Nuclear physics and lattice QCD}},
  \href{https://doi.org/10.22323/1.020.0020}{\emph{PoS} {\bfseries LAT2005}
  (2006) 020}.

\bibitem{Aoki:2007zf}
S.~Aoki, \emph{{Hadron interactions from lattice QCD}},
  \href{https://doi.org/10.22323/1.042.0002}{\emph{PoS} {\bfseries LATTICE2007}
  (2007) 002} [\href{https://arxiv.org/abs/0711.2151}{{\ttfamily 0711.2151}}].

\bibitem{Beane:2008ia}
S.~R. Beane, \emph{{Hadronic interactions and nuclear physics}},
  \href{https://doi.org/10.22323/1.066.0008}{\emph{PoS} {\bfseries LATTICE2008}
  (2008) 008} [\href{https://arxiv.org/abs/0812.1236}{{\ttfamily 0812.1236}}].

\bibitem{Detmold:2009zz}
W.~Detmold, \emph{{Multi-hadron systems in lattice QCD}},
  \href{https://doi.org/10.22323/1.091.0008}{\emph{PoS} {\bfseries LAT2009}
  (2009) 008}.

\bibitem{Hatsuda:2014kma}
T.~Hatsuda, \emph{{Nuclear physics on the lattice}},
  \href{https://doi.org/10.22323/1.105.0008}{\emph{PoS} {\bfseries LATTICE2010}
  (2010) 008}.

\bibitem{Orginos:2011zz}
K.~Orginos, \emph{{Hadron interactions}},
  \href{https://doi.org/10.22323/1.139.0016}{\emph{PoS} {\bfseries LATTICE2011}
  (2011) 016}.

\bibitem{Doi:2012ab}
{\scshape HAL QCD} collaboration, T.~Doi, \emph{{Nuclear physics from lattice
  simulations}}, \href{https://doi.org/10.22323/1.164.0009}{\emph{PoS}
  {\bfseries LATTICE2012} (2012) 009}
  [\href{https://arxiv.org/abs/1212.1572}{{\ttfamily 1212.1572}}].

\bibitem{Walker-Loud:2014iea}
A.~Walker-Loud, \emph{{Nuclear Physics Review}},
  \href{https://doi.org/10.22323/1.187.0013}{\emph{PoS} {\bfseries LATTICE2013}
  (2014) 013} [\href{https://arxiv.org/abs/1401.8259}{{\ttfamily 1401.8259}}].

\bibitem{Yamazaki:2015nka}
T.~Yamazaki, \emph{{Hadronic Interactions}},
  \href{https://doi.org/10.22323/1.214.0009}{\emph{PoS} {\bfseries LATTICE2014}
  (2015) 009} [\href{https://arxiv.org/abs/1503.08671}{{\ttfamily
  1503.08671}}].

\bibitem{Savage:2016egr}
M.~J. Savage, \emph{{Nuclear Physics}},
  \href{https://doi.org/10.22323/1.256.0021}{\emph{PoS} {\bfseries LATTICE2016}
  (2016) 021} [\href{https://arxiv.org/abs/1611.02078}{{\ttfamily
  1611.02078}}].

\bibitem{Davoudi:2017ddj}
Z.~Davoudi, \emph{{Lattice QCD input for nuclear structure and reactions}},
  \href{https://doi.org/10.1051/epjconf/201817501022}{\emph{EPJ Web Conf. 175}
  {\bfseries LATTICE2017} (2018) 01022}
  [\href{https://arxiv.org/abs/1711.02020}{{\ttfamily 1711.02020}}].

\bibitem{Beane:2006mx}
S.~Beane, P.~Bedaque, K.~Orginos and M.~Savage, \emph{{Nucleon-nucleon
  scattering from fully-dynamical lattice QCD}},
  \href{https://doi.org/10.1103/PhysRevLett.97.012001}{\emph{Phys.Rev.Lett.}
  {\bfseries 97} (2006) 012001}
  [\href{https://arxiv.org/abs/hep-lat/0602010}{{\ttfamily hep-lat/0602010}}].

\bibitem{Ishii:2006ec}
N.~Ishii, S.~Aoki and T.~Hatsuda, \emph{{The Nuclear Force from Lattice QCD}},
  \href{https://doi.org/10.1103/PhysRevLett.99.022001}{\emph{Phys.Rev.Lett.}
  {\bfseries 99} (2007) 022001}
  [\href{https://arxiv.org/abs/nucl-th/0611096}{{\ttfamily nucl-th/0611096}}].

\bibitem{Beane:2010hg}
{\scshape NPLQCD} collaboration, S.~Beane et~al., \emph{{Evidence for a Bound
  H-dibaryon from Lattice QCD}},
  \href{https://doi.org/10.1103/PhysRevLett.106.162001}{\emph{Phys.Rev.Lett.}
  {\bfseries 106} (2011) 162001}
  [\href{https://arxiv.org/abs/1012.3812}{{\ttfamily 1012.3812}}].

\bibitem{Inoue:2010hs}
{\scshape HAL QCD} collaboration, T.~Inoue et~al., \emph{{Baryon-Baryon
  Interactions in the Flavor SU(3) Limit from Full QCD Simulations on the
  Lattice}}, \href{https://doi.org/10.1143/PTP.124.591}{\emph{Prog.Theor.Phys.}
  {\bfseries 124} (2010) 591}
  [\href{https://arxiv.org/abs/1007.3559}{{\ttfamily 1007.3559}}].

\bibitem{Inoue:2010es}
{\scshape HAL QCD} collaboration, T.~Inoue et~al., \emph{{Bound H-dibaryon in
  Flavor SU(3) Limit of Lattice QCD}},
  \href{https://doi.org/10.1103/PhysRevLett.106.162002}{\emph{Phys.Rev.Lett.}
  {\bfseries 106} (2011) 162002}
  [\href{https://arxiv.org/abs/1012.5928}{{\ttfamily 1012.5928}}].

\bibitem{Nemura:2008sp}
H.~Nemura, N.~Ishii, S.~Aoki and T.~Hatsuda, \emph{{Hyperon-nucleon force from
  lattice QCD}},
  \href{https://doi.org/10.1016/j.physletb.2009.02.003}{\emph{Phys.Lett.}
  {\bfseries B673} (2009) 136}
  [\href{https://arxiv.org/abs/0806.1094}{{\ttfamily 0806.1094}}].

\bibitem{Yamazaki:2010ya}
{\scshape PACS-CS} collaboration, T.~Yamazaki, \emph{{Calculation of Helium
  nuclei in quenched lattice QCD}},
  \href{https://doi.org/10.22323/1.105.0021}{\emph{PoS} {\bfseries LATTICE2010}
  (2010) 021} [\href{https://arxiv.org/abs/1012.0410}{{\ttfamily 1012.0410}}].

\bibitem{Beane:2011iw}
{\scshape NPLQCD} collaboration, S.~Beane et~al., \emph{{The Deuteron and
  Exotic Two-Body Bound States from Lattice QCD}},
  \href{https://doi.org/10.1103/PhysRevD.85.054511}{\emph{Phys.Rev.} {\bfseries
  D85} (2012) 054511} [\href{https://arxiv.org/abs/1109.2889}{{\ttfamily
  1109.2889}}].

\bibitem{Doi:2011gq}
{\scshape HAL QCD} collaboration, T.~Doi, S.~Aoki, T.~Hatsuda, Y.~Ikeda,
  T.~Inoue, N.~Ishii et~al., \emph{{Exploring Three-Nucleon Forces in Lattice
  QCD}}, \href{https://doi.org/10.1143/PTP.127.723}{\emph{Prog. Theor. Phys.}
  {\bfseries 127} (2012) 723}
  [\href{https://arxiv.org/abs/1106.2276}{{\ttfamily 1106.2276}}].

\bibitem{Inoue:2011ai}
{\scshape HAL QCD} collaboration, T.~Inoue et~al., \emph{{Two-Baryon Potentials
  and H-Dibaryon from 3-flavor Lattice QCD Simulations}},
  \href{https://doi.org/10.1016/j.nuclphysa.2012.02.008}{\emph{Nucl.Phys.}
  {\bfseries A881} (2012) 28}
  [\href{https://arxiv.org/abs/1112.5926}{{\ttfamily 1112.5926}}].

\bibitem{Yamazaki:2011nd}
{\scshape PACS-CS} collaboration, T.~Yamazaki, Y.~Kuramashi and A.~Ukawa,
  \emph{{Two-Nucleon Bound States in Quenched Lattice QCD}},
  \href{https://doi.org/10.1103/PhysRevD.84.054506}{\emph{Phys.Rev.} {\bfseries
  D84} (2011) 054506} [\href{https://arxiv.org/abs/1105.1418}{{\ttfamily
  1105.1418}}].

\bibitem{Aoki:2012tk}
{\scshape HAL QCD} collaboration, S.~Aoki, T.~Doi, T.~Hatsuda, Y.~Ikeda,
  T.~Inoue, N.~Ishii et~al., \emph{{Lattice QCD approach to Nuclear Physics}},
  \href{https://doi.org/10.1093/ptep/pts010}{\emph{PTEP} {\bfseries 2012}
  (2012) 01A105} [\href{https://arxiv.org/abs/1206.5088}{{\ttfamily
  1206.5088}}].

\bibitem{Beane:2012ey}
S.~Beane, E.~Chang, S.~Cohen, W.~Detmold, H.-W. Lin et~al.,
  \emph{{Hyperon-Nucleon Interactions and the Composition of Dense Nuclear
  Matter from Quantum Chromodynamics}},
  \href{https://doi.org/10.1103/PhysRevLett.109.172001}{\emph{Phys.Rev.Lett.}
  {\bfseries 109} (2012) 172001}
  [\href{https://arxiv.org/abs/1204.3606}{{\ttfamily 1204.3606}}].

\bibitem{Beane:2012vq}
{\scshape NPLQCD} collaboration, S.~Beane et~al., \emph{{Light Nuclei and
  Hypernuclei from Quantum Chromodynamics in the Limit of SU(3) Flavor
  Symmetry}},
  \href{https://doi.org/10.1103/PhysRevD.87.034506}{\emph{Phys.Rev.} {\bfseries
  D87} (2013) 034506} [\href{https://arxiv.org/abs/1206.5219}{{\ttfamily
  1206.5219}}].

\bibitem{Yamazaki:2012hi}
T.~Yamazaki, K.-i. Ishikawa, Y.~Kuramashi and A.~Ukawa, \emph{{Helium nuclei,
  deuteron and dineutron in 2+1 flavor lattice QCD}},
  \href{https://doi.org/10.1103/PhysRevD.86.074514}{\emph{Phys.Rev.} {\bfseries
  D86} (2012) 074514} [\href{https://arxiv.org/abs/1207.4277}{{\ttfamily
  1207.4277}}].

\bibitem{Yamazaki:2012fn}
T.~Yamazaki, K.-i. Ishikawa, Y.~Kuramashi and A.~Ukawa, \emph{{Bound states of
  multi-nucleon channels in $N_f$=2+1 lattice QCD}},
  \href{https://doi.org/10.22323/1.164.0143}{\emph{PoS} {\bfseries LATTICE2012}
  (2012) 143} [\href{https://arxiv.org/abs/1211.4334}{{\ttfamily 1211.4334}}].

\bibitem{Beane:2013br}
{\scshape NPLQCD} collaboration, S.~Beane et~al., \emph{{Nucleon-Nucleon
  Scattering Parameters in the Limit of SU(3) Flavor Symmetry}},
  \href{https://doi.org/10.1103/PhysRevC.88.024003}{\emph{Phys.Rev.} {\bfseries
  C88} (2013) 024003} [\href{https://arxiv.org/abs/1301.5790}{{\ttfamily
  1301.5790}}].

\bibitem{Murano:2013xxa}
{\scshape HAL QCD} collaboration, K.~Murano et~al., \emph{{Spin--orbit force
  from lattice QCD}},
  \href{https://doi.org/10.1016/j.physletb.2014.05.061}{\emph{Phys.Lett.}
  {\bfseries B735} (2014) 19}
  [\href{https://arxiv.org/abs/1305.2293}{{\ttfamily 1305.2293}}].

\bibitem{Orginos:2015aya}
K.~Orginos, A.~Parreno, M.~J. Savage, S.~R. Beane, E.~Chang and W.~Detmold,
  \emph{{Two nucleon systems at $m_\pi\sim 450~{\rm MeV}$ from lattice QCD}},
  \href{https://doi.org/10.1103/PhysRevD.92.114512}{\emph{Phys. Rev.}
  {\bfseries D92} (2015) 114512}
  [\href{https://arxiv.org/abs/1508.07583}{{\ttfamily 1508.07583}}].

\bibitem{Yamazaki:2015asa}
T.~Yamazaki, K.-i. Ishikawa, Y.~Kuramashi and A.~Ukawa, \emph{{Study of quark
  mass dependence of binding energy for light nuclei in 2+1 flavor lattice
  QCD}}, \href{https://doi.org/10.1103/PhysRevD.92.014501}{\emph{Phys. Rev.}
  {\bfseries D92} (2015) 014501}
  [\href{https://arxiv.org/abs/1502.04182}{{\ttfamily 1502.04182}}].

\bibitem{Berkowitz:2015eaa}
E.~Berkowitz, T.~Kurth, A.~Nicholson, B.~Jo\'{o}, E.~Rinaldi, M.~Strother
  et~al., \emph{{Two-Nucleon Higher Partial-Wave Scattering from Lattice QCD}},
  \href{https://doi.org/10.1016/j.physletb.2016.12.024}{\emph{Phys. Lett.}
  {\bfseries B765} (2017) 285}
  [\href{https://arxiv.org/abs/1508.00886}{{\ttfamily 1508.00886}}].

\bibitem{Wagman:2017tmp}
M.~L. Wagman, F.~Winter, E.~Chang, Z.~Davoudi, W.~Detmold, K.~Orginos et~al.,
  \emph{{Baryon-Baryon Interactions and Spin-Flavor Symmetry from Lattice
  Quantum Chromodynamics}},
  \href{https://doi.org/10.1103/PhysRevD.96.114510}{\emph{Phys. Rev.}
  {\bfseries D96} (2017) 114510}
  [\href{https://arxiv.org/abs/1706.06550}{{\ttfamily 1706.06550}}].

\bibitem{luscher:1986n1}
M.~L\"uscher, \emph{Volume dependence of the energy spectrum in massive quantum
  field theories i}, {\emph{Communications in Mathematical Physics} {\bfseries
  104} (1986) 177}.

\bibitem{luscher:1986n2}
M.~Luscher, \emph{{Volume Dependence of the Energy Spectrum in Massive Quantum
  Field Theories. 2. Scattering States}},
  \href{https://doi.org/10.1007/BF01211097}{\emph{Commun.Math.Phys.} {\bfseries
  105} (1986) 153}.

\bibitem{Wiese:1989}
U.-J. Wiese, \emph{Identification of resonance parameters from the finite
  volume energy spectrum},
  \href{https://doi.org/10.1016/0920-5632(89)90171-0}{\emph{Nuclear Physics B -
  Proceedings Supplements} {\bfseries 9} (1989) 609 }.

\bibitem{Luscher:1991n1}
M.~Luscher, \emph{{Two particle states on a torus and their relation to the
  scattering matrix}},
  \href{https://doi.org/10.1016/0550-3213(91)90366-6}{\emph{Nucl.Phys.}
  {\bfseries B354} (1991) 531}.

\bibitem{Luscher:1991n2}
M.~L\"uscher, \emph{Signatures of unstable particles in finite volume},
  \href{https://doi.org/10.1016/0550-3213(91)90584-K}{\emph{Nuclear Physics B}
  {\bfseries 364} (1991) 237 }.

\bibitem{Rummukainen:1995vs}
K.~Rummukainen and S.~A. Gottlieb, \emph{{Resonance scattering phase shifts on
  a nonrest frame lattice}},
  \href{https://doi.org/10.1016/0550-3213(95)00313-H}{\emph{Nucl. Phys.}
  {\bfseries B450} (1995) 397}
  [\href{https://arxiv.org/abs/hep-lat/9503028}{{\ttfamily hep-lat/9503028}}].

\bibitem{Feng:2004ua}
X.~Feng, X.~Li and C.~Liu, \emph{{Two particle states in an asymmetric box and
  the elastic scattering phases}},
  \href{https://doi.org/10.1103/PhysRevD.70.014505}{\emph{Phys. Rev.}
  {\bfseries D70} (2004) 014505}
  [\href{https://arxiv.org/abs/hep-lat/0404001}{{\ttfamily hep-lat/0404001}}].

\bibitem{He:2005ey}
S.~He, X.~Feng and C.~Liu, \emph{{Two particle states and the S-matrix elements
  in multi-channel scattering}},
  \href{https://doi.org/10.1088/1126-6708/2005/07/011}{\emph{JHEP} {\bfseries
  07} (2005) 011} [\href{https://arxiv.org/abs/hep-lat/0504019}{{\ttfamily
  hep-lat/0504019}}].

\bibitem{Kim:2005gf}
C.~h. Kim, C.~T. Sachrajda and S.~R. Sharpe, \emph{{Finite-volume effects for
  two-hadron states in moving frames}},
  \href{https://doi.org/10.1016/j.nuclphysb.2005.08.029}{\emph{Nucl. Phys.}
  {\bfseries B727} (2005) 218}
  [\href{https://arxiv.org/abs/hep-lat/0507006}{{\ttfamily hep-lat/0507006}}].

\bibitem{Lage:2009zv}
M.~Lage, U.-G. Mei{\ss}ner and A.~Rusetsky, \emph{{A Method to measure the
  antikaon-nucleon scattering length in lattice QCD}},
  \href{https://doi.org/10.1016/j.physletb.2009.10.055}{\emph{Phys. Lett.}
  {\bfseries B681} (2009) 439}
  [\href{https://arxiv.org/abs/0905.0069}{{\ttfamily 0905.0069}}].

\bibitem{Bernard:2010fp}
V.~Bernard, M.~Lage, U.-G. Mei{\ss}ner and A.~Rusetsky, \emph{{Scalar mesons in
  a finite volume}}, \href{https://doi.org/10.1007/JHEP01(2011)019}{\emph{JHEP}
  {\bfseries 01} (2011) 019} [\href{https://arxiv.org/abs/1010.6018}{{\ttfamily
  1010.6018}}].

\bibitem{Fu:2011xz}
Z.~Fu, \emph{{Rummukainen-Gottlieb's formula on two-particle system with
  different mass}},
  \href{https://doi.org/10.1103/PhysRevD.85.014506}{\emph{Phys. Rev.}
  {\bfseries D85} (2012) 014506}
  [\href{https://arxiv.org/abs/1110.0319}{{\ttfamily 1110.0319}}].

\bibitem{Bernard:2012bi}
V.~Bernard, D.~Hoja, U.-G. Mei{\ss}ner and A.~Rusetsky, \emph{{Matrix elements
  of unstable states}},
  \href{https://doi.org/10.1007/JHEP09(2012)023}{\emph{JHEP} {\bfseries 09}
  (2012) 023} [\href{https://arxiv.org/abs/1205.4642}{{\ttfamily 1205.4642}}].

\bibitem{Briceno:2012yi}
R.~A. Brice{\~n}o and Z.~Davoudi, \emph{{Moving multichannel systems in a
  finite volume with application to proton-proton fusion}},
  \href{https://doi.org/10.1103/PhysRevD.88.094507}{\emph{Phys. Rev.}
  {\bfseries D88} (2013) 094507}
  [\href{https://arxiv.org/abs/1204.1110}{{\ttfamily 1204.1110}}].

\bibitem{Gockeler:2012yj}
M.~Gockeler, R.~Horsley, M.~Lage, U.-G. Mei{\ss}ner, P.~E.~L. Rakow,
  A.~Rusetsky et~al., \emph{{Scattering phases for meson and baryon resonances
  on general moving-frame lattices}},
  \href{https://doi.org/10.1103/PhysRevD.86.094513}{\emph{Phys. Rev.}
  {\bfseries D86} (2012) 094513}
  [\href{https://arxiv.org/abs/1206.4141}{{\ttfamily 1206.4141}}].

\bibitem{Guo:2012hv}
P.~Guo, J.~Dudek, R.~Edwards and A.~P. Szczepaniak, \emph{{Coupled-channel
  scattering on a torus}},
  \href{https://doi.org/10.1103/PhysRevD.88.014501}{\emph{Phys. Rev.}
  {\bfseries D88} (2013) 014501}
  [\href{https://arxiv.org/abs/1211.0929}{{\ttfamily 1211.0929}}].

\bibitem{Briceno:2014oea}
R.~A. Brice{\~n}o, \emph{{Two-particle multichannel systems in a finite volume
  with arbitrary spin}},
  \href{https://doi.org/10.1103/PhysRevD.89.074507}{\emph{Phys. Rev.}
  {\bfseries D89} (2014) 074507}
  [\href{https://arxiv.org/abs/1401.3312}{{\ttfamily 1401.3312}}].

\bibitem{Lee:2017igf}
F.~X. Lee and A.~Alexandru, \emph{{Scattering phaseshift formulas for mesons
  and baryons in elongated boxes}},
  \href{https://arxiv.org/abs/1706.00262}{{\ttfamily 1706.00262}}.

\bibitem{Briceno:2018bnl}
R.~A. Brice\~{n}o, \emph{{Resonances from lattice QCD}},
  \href{https://doi.org/10.1051/epjconf/201817501016}{\emph{EPJ Web Conf. 175}
  {\bfseries LATTICE2017} (2018) 01016}.

\bibitem{Brett:2017yhm}
R.~Brett, J.~Bulava, J.~Fallica, A.~Hanlon, B.~H{\"o}rz, C.~Morningstar et~al.,
  \emph{{Scattering from finite-volume energies including higher partial waves
  and multiple decay channels}},
  \href{https://doi.org/10.1051/epjconf/201817505005}{\emph{EPJ Web Conf. 175}
  {\bfseries LATTICE2017} (2018) 05005}
  [\href{https://arxiv.org/abs/1710.04169}{{\ttfamily 1710.04169}}].

\bibitem{Bulava:2017stw}
J.~Bulava, B.~H{\"o}rz and C.~Morningstar, \emph{{Multi-hadron spectroscopy in
  a large physical volume}},
  \href{https://doi.org/10.1051/epjconf/201817505026}{\emph{EPJ Web Conf. 175}
  {\bfseries LATTICE2017} (2018) 05026}
  [\href{https://arxiv.org/abs/1710.04545}{{\ttfamily 1710.04545}}].

\bibitem{Wilson:2016rid}
D.~J. Wilson, \emph{{Resonances in Coupled-Channel Scattering}},
  \href{https://doi.org/10.22323/1.256.0016}{\emph{PoS} {\bfseries LATTICE2016}
  (2016) 016} [\href{https://arxiv.org/abs/1611.07281}{{\ttfamily
  1611.07281}}].

\bibitem{Briceno:2013hya}
R.~A. Brice{\~n}o, Z.~Davoudi, T.~C. Luu and M.~J. Savage, \emph{{Two-Baryon
  Systems with Twisted Boundary Conditions}},
  \href{https://doi.org/10.1103/PhysRevD.89.074509}{\emph{Phys. Rev.}
  {\bfseries D89} (2014) 074509}
  [\href{https://arxiv.org/abs/1311.7686}{{\ttfamily 1311.7686}}].

\bibitem{Korber:2015rce}
C.~K{\"o}rber and T.~Luu, \emph{{Applying Twisted Boundary Conditions for
  Few-body Nuclear Systems}},
  \href{https://doi.org/10.1103/PhysRevC.93.054002}{\emph{Phys. Rev.}
  {\bfseries C93} (2016) 054002}
  [\href{https://arxiv.org/abs/1511.06598}{{\ttfamily 1511.06598}}].

\bibitem{Hansen:2013dla}
M.~T. Hansen and S.~R. Sharpe, \emph{{Relativistic, model-independent,
  three-particle quantization condition}},
  \href{https://doi.org/10.22323/1.187.0221}{\emph{PoS} {\bfseries LATTICE2013}
  (2014) 221} [\href{https://arxiv.org/abs/1311.4848}{{\ttfamily 1311.4848}}].

\bibitem{Hansen:2014lya}
M.~T. Hansen and S.~R. Sharpe, \emph{{Relativistic three-particle quantization
  condition: an update}}, \href{https://doi.org/10.22323/1.214.0088}{\emph{PoS}
  {\bfseries LATTICE2014} (2015) 088}
  [\href{https://arxiv.org/abs/1409.7012}{{\ttfamily 1409.7012}}].

\bibitem{Hansen:2015azg}
M.~T. Hansen, \emph{{Extracting three-body observables from finite-volume
  quantities}}, \href{https://doi.org/10.22323/1.251.0008}{\emph{PoS}
  {\bfseries LATTICE2015} (2016) 008}
  [\href{https://arxiv.org/abs/1511.04737}{{\ttfamily 1511.04737}}].

\bibitem{Briceno:2016ffu}
R.~A. Brice{\~n}o, M.~T. Hansen and S.~R. Sharpe, \emph{{Progress on the
  three-particle quantization condition}},
  \href{https://doi.org/10.22323/1.256.0115}{\emph{PoS} {\bfseries LATTICE2016}
  (2016) 115} [\href{https://arxiv.org/abs/1609.09805}{{\ttfamily
  1609.09805}}].

\bibitem{sharpe:2018}
S.~Sharpe et~al., \emph{{Progress on relativistic three-particle quantization
  condition}}, {\emph{PoS} {\bfseries LATTICE2018} (2018) 076}.

\bibitem{mai:2018}
M.~Mai and M.~Doring, \emph{{3-body quantization condition in unitary isobar
  formalsim}}, {\emph{PoS} {\bfseries LATTICE2018} (2018) 050}.

\bibitem{ma:2018}
Y.~Ma and Y.~Chen, \emph{{Bethe-Salpeter wavefunctions of hybrid charmonia}},
  {\emph{PoS} {\bfseries LATTICE2018} (2018) 069}.

\bibitem{yamazaki:2018}
T.~Yamazaki and Y.~Kuramashi, \emph{{Relations between scattering amplitude and
  Bethe-Salpeter wave function in quantum field theory}}, {\emph{PoS}
  {\bfseries LATTICE2018} (2018) 077}.

\bibitem{namekawa:2018}
Y.~Namekawa and T.~Yamazaki, \emph{{Scattering length from BS wave function
  inside the interaction range}}, {\emph{PoS} {\bfseries LATTICE2018} (2018)
  078}.

\bibitem{iritani:2018}
T.~Iritani, \emph{{HAL QCD method and Nucleon-Omega interaction with physical
  quark masses}}, {\emph{PoS} {\bfseries LATTICE2018} (2018) 090}.

\bibitem{doi:2018}
T.~Doi, \emph{{Baryon interactions at physical quark masses in Lattice QCD}},
  {\emph{PoS} {\bfseries LATTICE2018} (2018) 091}.

\bibitem{guo:2018}
Z.~Guo, \emph{{Confront the lattice finite-volume energy levels with chiral
  effective field theory}}, {\emph{PoS} {\bfseries LATTICE2018} (2018) 055}.

\bibitem{Wagman:2017jva}
M.~L. Wagman, \emph{{Life Outside the Golden Window: Statistical Angles on the
  Signal-to-Noise Problem}},
  \href{https://doi.org/10.1051/epjconf/201817505003}{\emph{EPJ Web Conf. 175}
  {\bfseries LATTICE2017} (2018) 05003}
  [\href{https://arxiv.org/abs/1710.10818}{{\ttfamily 1710.10818}}].

\bibitem{Basak:2005ir}
{\scshape Lattice Hadron Physics (LHPC)} collaboration, S.~Basak, R.~Edwards,
  G.~T. Fleming, U.~M. Heller, C.~Morningstar, D.~Richards et~al.,
  \emph{{Clebsch-Gordan construction of lattice interpolating fields for
  excited baryons}},
  \href{https://doi.org/10.1103/PhysRevD.72.074501}{\emph{Phys. Rev.}
  {\bfseries D72} (2005) 074501}
  [\href{https://arxiv.org/abs/hep-lat/0508018}{{\ttfamily hep-lat/0508018}}].

\bibitem{Beane:2009kya}
S.~R. Beane, W.~Detmold, T.~C. Luu, K.~Orginos, A.~Parreno, M.~J. Savage
  et~al., \emph{{High Statistics Analysis using Anisotropic Clover Lattices:
  (I) Single Hadron Correlation Functions}},
  \href{https://doi.org/10.1103/PhysRevD.79.114502}{\emph{Phys. Rev.}
  {\bfseries D79} (2009) 114502}
  [\href{https://arxiv.org/abs/0903.2990}{{\ttfamily 0903.2990}}].

\bibitem{Orginos:2014pma}
K.~Orginos, \emph{{Construction and Analysis of Two Baryon Correlation
  functions}}, \href{https://doi.org/10.22323/1.105.0118}{\emph{PoS} {\bfseries
  LATTICE2010} (2010) 118}.

\bibitem{Berkowitz:2017smo}
E.~Berkowitz, A.~Nicholson, C.~C. Chang, E.~Rinaldi, M.~A. Clark, B.~Jo\'{o}
  et~al., \emph{{Calm Multi-Baryon Operators}},
  \href{https://doi.org/10.1051/epjconf/201817505029}{\emph{EPJ Web Conf. 175}
  {\bfseries LATTICE2017} (2018) 05029}
  [\href{https://arxiv.org/abs/1710.05642}{{\ttfamily 1710.05642}}].

\bibitem{Doi:2012xd}
T.~Doi and M.~G. Endres, \emph{{Unified contraction algorithm for multi-baryon
  correlators on the lattice}},
  \href{https://doi.org/10.1016/j.cpc.2012.09.004}{\emph{Comput. Phys. Commun.}
  {\bfseries 184} (2013) 117}
  [\href{https://arxiv.org/abs/1205.0585}{{\ttfamily 1205.0585}}].

\bibitem{Detmold:2012eu}
W.~Detmold and K.~Orginos, \emph{{Nuclear correlation functions in lattice
  QCD}}, \href{https://doi.org/10.1103/PhysRevD.87.114512}{\emph{Phys. Rev.}
  {\bfseries D87} (2013) 114512}
  [\href{https://arxiv.org/abs/1207.1452}{{\ttfamily 1207.1452}}].

\bibitem{Orginos:2012js}
K.~Orginos and W.~Detmold, \emph{{Multi-baryon systems}},
  \href{https://doi.org/10.22323/1.164.0147}{\emph{PoS} {\bfseries LATTICE2012}
  (2012) 147}.

\bibitem{Vachaspati:2014bda}
P.~Vachaspati and W.~Detmold, \emph{{Fast Evaluation of Multi-Hadron
  Correlation Functions}},
  \href{https://doi.org/10.22323/1.214.0041}{\emph{PoS} {\bfseries LATTICE2014}
  (2015) 041} [\href{https://arxiv.org/abs/1411.3691}{{\ttfamily 1411.3691}}].

\bibitem{Francis:2018qch}
A.~Francis, J.~R. Green, P.~M. Junnarkar, C.~Miao, T.~D. Rae and H.~Wittig,
  \emph{{Lattice QCD study of the $H$ dibaryon using hexaquark and two-baryon
  interpolators}},  \href{https://arxiv.org/abs/1805.03966}{{\ttfamily
  1805.03966}}.

\bibitem{Aoki:2017byw}
S.~Aoki, T.~Doi and T.~Iritani, \emph{{Sanity check for $NN$ bound states in
  lattice QCD with L\"{u}scher’s finite volume formula -- Disclosing Symptoms
  of Fake Plateaux}},
  \href{https://doi.org/10.1051/epjconf/201817505006}{\emph{EPJ Web Conf. 175}
  {\bfseries LATTICE2017} (2018) 05006}
  [\href{https://arxiv.org/abs/1707.08800}{{\ttfamily 1707.08800}}].

\bibitem{Luu:2011ep}
T.~Luu and M.~J. Savage, \emph{{Extracting Scattering Phase-Shifts in Higher
  Partial-Waves from Lattice QCD Calculations}},
  \href{https://doi.org/10.1103/PhysRevD.83.114508}{\emph{Phys. Rev.}
  {\bfseries D83} (2011) 114508}
  [\href{https://arxiv.org/abs/1101.3347}{{\ttfamily 1101.3347}}].

\bibitem{Nicholson:2015pys}
A.~Nicholson, E.~Berkowitz, E.~Rinaldi, P.~Vranas, T.~Kurth, B.~Jo\'{o} et~al.,
  \emph{{Two-nucleon scattering in multiple partial waves}},
  \href{https://doi.org/10.22323/1.251.0083}{\emph{PoS} {\bfseries LATTICE2015}
  (2016) 083} [\href{https://arxiv.org/abs/1511.02262}{{\ttfamily
  1511.02262}}].

\bibitem{Bazavov:2009jc}
{\scshape MILC} collaboration, A.~Bazavov et~al., \emph{{HISQ action in
  dynamical simulations}},
  \href{https://doi.org/10.22323/1.066.0033}{\emph{PoS} {\bfseries LATTICE2008}
  (2008) 033} [\href{https://arxiv.org/abs/0903.0874}{{\ttfamily 0903.0874}}].

\bibitem{Bazavov:2009wm}
{\scshape MILC} collaboration, A.~Bazavov et~al., \emph{{Progress on four
  flavor QCD with the HISQ action}},
  \href{https://doi.org/10.22323/1.091.0123}{\emph{PoS} {\bfseries LAT2009}
  (2009) 123} [\href{https://arxiv.org/abs/0911.0869}{{\ttfamily 0911.0869}}].

\bibitem{Bazavov:2010pi}
A.~Bazavov et~al., \emph{{Simulations with dynamical HISQ quarks}},
  \href{https://doi.org/10.22323/1.105.0320}{\emph{PoS} {\bfseries LATTICE2010}
  (2010) 320} [\href{https://arxiv.org/abs/1012.1265}{{\ttfamily 1012.1265}}].

\bibitem{Bazavov:2012xda}
{\scshape MILC} collaboration, A.~Bazavov et~al., \emph{{Lattice QCD ensembles
  with four flavors of highly improved staggered quarks}},
  \href{https://doi.org/10.1103/PhysRevD.87.054505}{\emph{Phys. Rev.}
  {\bfseries D87} (2013) 054505}
  [\href{https://arxiv.org/abs/1212.4768}{{\ttfamily 1212.4768}}].

\bibitem{Berkowitz:2017opd}
E.~Berkowitz, C.~Bouchard, C.~Chang, M.~Clark, B.~Jo\'{o}, T.~Kurth et~al.,
  \emph{{M\"obius Domain-Wall fermions on gradient-flowed dynamical HISQ
  ensembles}}, \href{https://doi.org/10.1103/PhysRevD.96.054513}{\emph{Phys.
  Rev.} {\bfseries D96} (2017) 054513}
  [\href{https://arxiv.org/abs/1701.07559}{{\ttfamily 1701.07559}}].

\bibitem{Nicholson:2016byl}
A.~Nicholson, E.~Berkowitz, C.~C. Chang, M.~A. Clark, B.~Jo\'{o}, T.~Kurth
  et~al., \emph{{Neutrinoless double beta decay from lattice QCD}},
  \href{https://doi.org/10.22323/1.256.0017}{\emph{PoS} {\bfseries LATTICE2016}
  (2016) 017} [\href{https://arxiv.org/abs/1608.04793}{{\ttfamily
  1608.04793}}].

\bibitem{monge-camacho:2018}
H.~Monge-Camacho et~al., \emph{{Short Range Operator Contributions to
  neutrinoless double beta decay from LQCD}}, {\emph{PoS} {\bfseries
  LATTICE2018} (2018) {263}}.

\bibitem{Chang:2017oll}
C.~C. Chang et~al., \emph{{Nucleon axial coupling from Lattice QCD}},
  \href{https://doi.org/10.1051/epjconf/201817501008}{\emph{EPJ Web Conf. 175}
  {\bfseries LATTICE2017} (2018) 01008}
  [\href{https://arxiv.org/abs/1710.06523}{{\ttfamily 1710.06523}}].

\bibitem{Chang:2018uxx}
C.~C. Chang et~al., \emph{{A per-cent-level determination of the nucleon axial
  coupling from quantum chromodynamics}},
  \href{https://doi.org/10.1038/s41586-018-0161-8}{\emph{Nature} {\bfseries
  558} (2018) 91} [\href{https://arxiv.org/abs/1805.12130}{{\ttfamily
  1805.12130}}].

\bibitem{sallmen:2018}
K.~Sallmen et~al., \emph{{Exploring the convergence of SU(2) HB$\chi$PT}},
  {\emph{PoS} {\bfseries LATTICE2018} (2018) 047}.

\bibitem{walker-loud:2018}
A.~Walker-Loud et~al., \emph{{Lattice QCD Spectroscopy for hadronic CP
  violation}}, {\emph{PoS} {\bfseries LATTICE2018} (2018) {066}}.

\bibitem{chang:2018}
C.~C. Chang et~al., \emph{{The Slope of Form Factors from Lattice QCD}},
  {\emph{PoS} {\bfseries LATTICE2018} (2018) {051}}.

\bibitem{gambhir:2018}
A.~Gambhir et~al., \emph{{Improving the Feynman-Hellman Method}}, {\emph{PoS}
  {\bfseries LATTICE2018} (2018) {126}}.

\bibitem{carpenter:2018}
L.~Carpenter et~al., \emph{{Scale Setting on the MDWF in Gradient Flow HISQ
  Action with the Omega Baryon}}, {\emph{PoS} {\bfseries LATTICE2018} (2018)
  222}.

\bibitem{Renner:2004ck}
{\scshape LHP} collaboration, D.~B. Renner, W.~Schroers, R.~Edwards, G.~T.
  Fleming, P.~Hagler, J.~W. Negele et~al., \emph{{Hadronic physics with
  domain-wall valence and improved staggered sea quarks}},
  \href{https://doi.org/10.1016/j.nuclphysbps.2004.11.357}{\emph{Nucl. Phys.
  Proc. Suppl.} {\bfseries 140} (2005) 255}
  [\href{https://arxiv.org/abs/hep-lat/0409130}{{\ttfamily hep-lat/0409130}}].

\bibitem{Bar:2005tu}
O.~Bar, C.~Bernard, G.~Rupak and N.~Shoresh, \emph{{Chiral perturbation theory
  for staggered sea quarks and Ginsparg-Wilson valence quarks}},
  \href{https://doi.org/10.1103/PhysRevD.72.054502}{\emph{Phys. Rev.}
  {\bfseries D72} (2005) 054502}
  [\href{https://arxiv.org/abs/hep-lat/0503009}{{\ttfamily hep-lat/0503009}}].

\bibitem{Tiburzi:2005is}
B.~C. Tiburzi, \emph{{Baryons with Ginsparg-Wilson quarks in a staggered sea}},
  \href{https://doi.org/10.1103/PhysRevD.72.094501,
  10.1103/PhysRevD.79.039904}{\emph{Phys. Rev.} {\bfseries D72} (2005) 094501}
  [\href{https://arxiv.org/abs/hep-lat/0508019}{{\ttfamily hep-lat/0508019}}].

\bibitem{Jiang:2007sn}
F.-J. Jiang, \emph{{Mixed Action Lattice Spacing Effects on the Nucleon Axial
  Charge}},  \href{https://arxiv.org/abs/hep-lat/0703012}{{\ttfamily
  hep-lat/0703012}}.

\bibitem{Kim:2014mpa}
H.-J. Kim and T.~Izubuchi, \emph{{M\"{o}bius domain wall fermion method on
  QUDA}}, \href{https://doi.org/10.22323/1.187.0033}{\emph{PoS} {\bfseries
  LATTICE2013} (2014) 033}.

\bibitem{Yamazaki:2013rna}
T.~Yamazaki, K.-I. Ishikawa, Y.~Kuramashi and A.~Ukawa, \emph{{Multi-nucleon
  bound states in $N_f=2+1$ lattice QCD}},
  \href{https://doi.org/10.22323/1.187.0230}{\emph{PoS} {\bfseries LATTICE2013}
  (2014) 230} [\href{https://arxiv.org/abs/1310.5797}{{\ttfamily 1310.5797}}].

\bibitem{Yamazaki:2015vjn}
{\scshape PACS} collaboration, T.~Yamazaki, \emph{{Light Nuclei and Nucleon
  Form Factors in $N_f=2+1$ Lattice QCD}},
  \href{https://doi.org/10.22323/1.251.0081}{\emph{PoS} {\bfseries LATTICE2015}
  (2016) 081} [\href{https://arxiv.org/abs/1511.09179}{{\ttfamily
  1511.09179}}].

\bibitem{wynen:2018}
J.-L. Wynen, E.~Berkowitz, T.~Luu, A.~Shindler and J.~Bulava, \emph{{Three
  neutrons from Lattice QCD}}, {\emph{PoS} {\bfseries LATTICE2018} (2018)
  {092}}.

\bibitem{Luu:2008fg}
T.~Luu, \emph{{Three fermions in a box}},
  \href{https://doi.org/10.22323/1.066.0246}{\emph{PoS} {\bfseries LATTICE2008}
  (2008) 246} [\href{https://arxiv.org/abs/0810.2331}{{\ttfamily 0810.2331}}].

\bibitem{Barnea:2018}
Barnea, Eliyahu and Bazak, \emph{{Properties of light lattice nuclei from
  EFT}}, .

\bibitem{McElvain:2016zbm}
K.~S. McElvain and W.~C. Haxton, \emph{{Nuclear Physics without High-Momentum
  Potentials: Direct Construction of the Effective Interaction from Scattering
  Observables}},  \href{https://arxiv.org/abs/1607.06863}{{\ttfamily
  1607.06863}}.

\bibitem{McElvain:2017}
K.~S. McElvain, \emph{{Construction of the Nuclear Effective Interaction from
  Energy Eigenstates and Boundary Conditions}}, {\emph{APS Meeting} (2017)
  \href{https://meetings.aps.org/Meeting/APR17/Session/C13.5}{BAPS.2017.APR.C13.5}}.

\bibitem{hanlon:2018}
A.~Hanlon, \emph{{Progress towards understanding the H-dibaryon from lattice
  QCD}}, {\emph{PoS} {\bfseries LATTICE2018} (2018) 081}.

\bibitem{Edwards:2004sx}
{\scshape SciDAC, LHPC, UKQCD} collaboration, R.~G. Edwards and B.~Jo\'{o},
  \emph{{The Chroma software system for lattice QCD}},
  \href{https://doi.org/10.1016/j.nuclphysbps.2004.11.254}{\emph{Nucl.Phys.Proc.Suppl.}
  {\bfseries 140} (2005) 832}
  [\href{https://arxiv.org/abs/hep-lat/0409003}{{\ttfamily hep-lat/0409003}}].

\bibitem{Clark:2009wm}
M.~Clark, R.~Babich, K.~Barros, R.~Brower and C.~Rebbi, \emph{{Solving Lattice
  QCD systems of equations using mixed precision solvers on GPUs}},
  \href{https://doi.org/10.1016/j.cpc.2010.05.002}{\emph{Comput.Phys.Commun.}
  {\bfseries 181} (2010) 1517}
  [\href{https://arxiv.org/abs/0911.3191}{{\ttfamily 0911.3191}}].

\bibitem{Babich:2011np}
R.~Babich, M.~Clark, B.~Jo\'{o}, G.~Shi, R.~Brower et~al., \emph{{Scaling
  Lattice QCD beyond 100 GPUs}},
  \href{https://arxiv.org/abs/1109.2935}{{\ttfamily 1109.2935}}.

\bibitem{hdf5}
{The HDF Group}, \emph{{Hierarchical Data Format, version 5}},  1997-NNNN.

\bibitem{Kurth:2015mqa}
T.~Kurth, A.~Pochinsky, A.~Sarje, S.~Syritsyn and A.~Walker-Loud,
  \emph{{High-Performance I/O: HDF5 for Lattice QCD}}, {\emph{PoS} {\bfseries
  LATTICE2014} (2015) 045} [\href{https://arxiv.org/abs/1501.06992}{{\ttfamily
  1501.06992}}].

\bibitem{Berkowitz:2017vcp}
E.~Berkowitz, \emph{{\texttt{METAQ}: Bundle Supercomputing Tasks}},
  \href{https://arxiv.org/abs/1702.06122}{{\ttfamily 1702.06122}}.

\bibitem{Berkowitz:2017xna}
E.~Berkowitz, G.~R. Jansen, K.~McElvain and A.~Walker-Loud, \emph{{Job
  Management and Task Bundling}},
  \href{https://doi.org/10.1051/epjconf/201817509007}{\emph{EPJ Web Conf. 175}
  {\bfseries LATTICE2017} (2018) 09007}
  [\href{https://arxiv.org/abs/1710.01986}{{\ttfamily 1710.01986}}].

\end{thebibliography}\endgroup
\end{multicols}

\end{document}